\documentclass[12pt,oneside]{article}
\usepackage{latexsym, amsmath, amscd, amssymb, theorem, epsfig}

\DeclareFontFamily{OT1}{rsfs10}{}
\DeclareFontShape{OT1}{rsfs10}{m}{n}{ <-> rsfs10 }{}
\DeclareMathAlphabet{\mathscript}{OT1}{rsfs10}{m}{n}

\numberwithin{equation}{section}

\newcommand{\rfe}[1]{(\ref{eq:#1})}

\newcommand{\I}{I}
\newcommand{\IIa}{IIa}
\newcommand{\IIb}{IIb}
\newcommand{\CY}[1]{CY$_#1$}
\newcommand{\tr}{{\rm tr}~}

\newcommand{\Hom}{{\rm Hom}}

\newcommand{\CO}{{\cal O}}

\newcommand{\CG}{{\cal G}}
\newcommand{\pasl}{\pa\kern-.55em /}

\newcommand{\ksl}{k\kern-.55em /}

\newcommand{\p}{{\partial}}
\newcommand{\ad}{{\rm ad~}}
\newcommand{\oprod}{\prod}

\DeclareFixedFont{\xiiss}{OT1}{cmss}{m}{n}{12}
\DeclareFixedFont{\ixss}{OT1}{cmss}{m}{n}{9}
\DeclareFixedFont{\cmrnine}{OT1}{cmr}{m}{n}{9}
\newcommand{\field}[1]{\mathbb{#1}}
\newcommand{\BC}{{\field C}}
\newcommand{\BR}{{\field R}}
\newcommand{\BZ}{{\field Z}}

\newcommand{\CP}{{\field C P}}
\newcommand{\CCs}{\hbox{\ixss C\kern-.4emI}}
\newcommand{\ZZs}{\hbox{\ixss Z\kern-.4emZ}}

\newcommand{\CN}{{\cal N}}

\makeatletter
\def\@strike{\relax\leavevmode
  \ifmmode
    \expandafter\mathpalette\expandafter\math@strike
  \else
    \expandafter\make@strike
  \fi}
\def\math@strike#1#2{%
  \setbox\z@\hbox{$\m@th#1{#2}$}\fin@strike}
\def\make@strike#1{%
  \setbox\z@\hbox{\color@begingroup#1\color@endgroup}\fin@strike}
\def\fin@strike{%
  \@tempdima\dp\z@
  \lower\@tempdima\hbox{\strike@start}%
  \box\z@
  \raise\@tempdimb\hbox{\strike@end}}
\def\strike@start{\special{ps: %
    currentpoint /starty exch def /startx exch def}}
\def\strike@end{
}
\newcommand\fs{\protect\@strike}

\begin{document}


\begin{titlepage}

\begin{center}
\hfill hep-th/0403018\\
\hfill RUNHETC-2004-05\\
\hfill UK/04-04

\vskip 1.5 cm
{\huge \bf Chirality Change in String Theory}
\vskip 1.3 cm
{\large Miclael R. Douglas$^{1,2,3}$ and Chengang Zhou$^4$}\\
\vskip 0.5 cm
{$^1$ Department of Physics and Astronomy \\
Rutgers University \\
Piscataway, NJ 08855-0849, USA\\
$^2$
I.H.E.S., Le Bois-Marie, Bures-sur-Yvette, 91440 France\\
$^3$
Caltech 452-48, Pasadena, CA 91125, USA\\
$^4$
Department of Astronomy and Physics \\
University of Kentucky\\
Lexington, KY 40506, USA}
\vskip 0.3cm
{mrd@physics.rutgers.edu\\
czhou@pa.uky.edu}
\end{center}

\date{}

\vskip 0.5 cm
\begin{abstract}

It is known that string theory compactifications leading to low energy
effective theories with different chiral matter content ({\it e.g.}
different numbers of standard model generations) are connected through
phase transitions, described by non-trivial quantum fixed point
theories.

We point out that such compactifications are also connected on a
purely classical level, through transitions that can be described
using standard effective field theory.  We illustrate this with
examples, including some in which the transition proceeds entirely
through supersymmetric configurations.

\end{abstract}

\thispagestyle{empty}

\end{titlepage}


\tableofcontents
\vfill\eject

\section{Introduction}

If string theory is truly the right unified theory, it has to make
contact with the real world by reproducing the Standard Model
at low energies.  One of the signatures of the Standard Model is the
existence of three generations of leptons and quarks, and one of
the early successes of string theory was a natural mechanism
producing a multiplicity of generations minus antigenerations, in
the context of heterotic string compactification on a Calabi-Yau
threefold (\CY3) $M$, using simple topological properties of the
\CY3 \cite{CHSW}.

Subsequent work led to many generalizations and other
constructions leading to grand unified theories with $\CN=1$
supersymmetry at a low scale, not to mention more radical
alternatives.  In many ways the simplest generalization is to keep
as the starting point $d=10$, $\CN=1$ supergravity/super
Yang-Mills with $E_8\times E_8$ or $SO(32)$ gauge group and with
the stringy corrections needed for anomaly cancellation, and still
compactify on a \CY3, only generalizing by taking a general
background gauge connection for a vector bundle $V$ satisfying
anomaly cancellation constraints \cite{HW, Wittenissues}. These
are often referred to as ``$(0,2)$'' models from the world-sheet
supersymmetry of the heterotic sigma model; in nonperturbative
heterotic string theory this class of models is further
generalized \cite{HoravaWitten}, but the number of generations
minus antigenerations, which we will denote $N_{gen}$, is still
determined by the choice of $V$.  Most of the other constructions,
including type \IIa\ and \IIb\ orientifold constructions, are dual
to these two constructions in a fairly clear way.

This picture raises many questions, such as: for which choices of
CY $M$ does there exist a three generation model ?  How do we find
its low energy Lagrangian ?  Which choices lead to stable vacua
with low scale $\CN=1$ supersymmetry ?  If there are many, which
additional criteria if any pick out the physically relevant one ?
Somewhat simpler but still relevant questions are, for a given CY
$M$, what are possible values of $N_{gen}$, and are vacua with
different $N_{gen}$ connected in the configuration space of string
theory?

In this paper we discuss whether vacua with different chiral
matter spectrum, including standard models with different numbers
of generations, are connected in string theory.  Previous work on
this question includes that of Kachru and Silverstein
\cite{KS} who discuss a mechanism which
connects such vacua through a non-trivial phase transition, whose
IR behavior is governed by the dimensional reduction of the phase
transition associated to the $(1,0)$ superconformal theory in six
dimensions. Further work supporting this idea appears in
\cite{HetMInstanton}.

One motivation for the present work is to refute a (still) common
belief that the number of generations provides an ``index'' which in
classical physics is conserved under continuous variations of the
background, so that such transitions could only proceed through a
phase transition.

This is false and in fact there is a very simple argument that all
vacua obtained by making different choices of $V$ must be
connected already on the classical level, without going through
non-classical phase transitions.  We will give this in more detail
below, but the essence in the heterotic string construction is
just the following.  We recall that the low energy gauge group is
the commutant in the ten dimensional gauge group (say) $E_8\times
E_8$ of the structure group $G$ of the bundle $V$.  To get a
realistic grand unified gauge group, we take $V$ to have structure
group $SU(k)$ (for example $k=4$ leads to unbroken $SO(10)$).  The
number of generations is then $N_{gen}=c_3(V)/2$, as follows from
the index theorem \cite{GSW}.

Now for bundles with structure group $SU(k)$, $c_3(V)$ is a
topological invariant.  On the other hand, if we allow general
motions in configuration space (possibly through a potential
barrier), nothing prevents us from deforming the connection on $V$
into the full ten dimensional gauge group.  Now, for $E_8$ and
$Spin(32)/\BZ_2$ bundles, $c_3(V)$ is not a topological invariant;
in fact it cannot even be defined: its definition uses the third
order symmetric invariant $d^{abc}$, which does not exist for
these Lie algebras.  Indeed, as is familiar from other discussions
\cite{DMW}, all $E_8$ vector bundles on a manifold $M$ of (real)
dimension less than $15$ with a specified rank and second Chern
class $c_2(V)$ are connected by smooth deformations.  Such a
deformation will correspond to a path in the configuration space
of the full string theory, connecting vacua with two possibly
different values of $N_{gen}$.

Naively, this might be thought paradoxical as chiral matter cannot
be lifted by varying uncharged fields or parameters in a low
energy effective Lagrangian.  The obvious loophole which this
suggests is that chiral matter must be lifted by giving vevs to
charged fields, as is familiar even in the standard model.  When
the gauge connection in $V$ deforms into $E_8$ (or any larger
group containing the original $SU(k)$), some or all of the low
energy gauge symmetry is broken.  If the chiral matter is
nonchiral under the remaining unbroken gauge symmetry, there is no
barrier to giving it a mass.  As an example, after a deformation
breaking $SO(10)$ to $SO(9)$, a chiral spinor in the $16$ of
$SO(10)$ becomes a nonchiral spinor of $SO(9)$, which can be
lifted.

The remaining question is how the chiral matter can remain massive
when the original gauge symmetry is restored with different $c_3$.
The answer to this question is that the final result is a new
model in which the original gauge symmetry is {\it not} restored.
Rather, an isomorphic but different copy of the gauge group
becomes unbroken. One should think of the low energy gauge group
as a small unbroken subgroup of a large (even infinite
dimensional) gauge group, the group of all gauge transformations
of $E_8$ gauge theory in six dimensions, almost all of which is
broken by the background.  Even if the unbroken gauge groups $G$
are isomorphic on the two sides of the transition, they can be
{\it two different embeddings} of this gauge group in the large
gauge group.

The minimal effective field theory realization of this phenomenon
is in terms of $G\times G$ gauge theory, with matter chosen to
always break at least one copy of $G$, and along the interpolating
path break to a subgroup of $G$.  We will exhibit such effective
Lagrangians in section 2.

These simple arguments more or less suffice for a basic physical
understanding of the phenomenon.  Most of the paper is devoted to
seeing how this works in string theory examples, using fairly
elementary methods, and dispelling various paradoxes which arise.

We also consider some more detailed questions.  One of these is to ask
what is the potential barrier to such a transition, a question which
might enter in early cosmology or in using some hypothetical ``vacuum
selection principle.''  At present there is no good proposal for such
a principle, and at present we do not have good control over
nonperturbative corrections to the potential, but the question is
still interesting to consider.

One expects the mechanism of \cite{KS} to involve no potential
barrier, because it goes through a superconformal theory, with no
preferred energy scale.  As for the classical mechanism we
discuss, on general grounds, one might expect variations of the
gauge connection on a compact manifold of ``size'' $R$ to go
through a potential barrier of height $V \sim 1/g_{YM4}^2 R^4$.
This can be made small at large radius or strong coupling but
would be non-zero.

Is it possible that examples exist with no potential barrier?  To
get at this question, we discuss explicit string theory examples.
We start in section 2 with toy models, gauge theory compactified
on $S^1$ and $S^2$, which are easy to work out in full detail. The
infinite dimensional gauge groups which arise here are the loop
groups $\hat G$.  One already finds on $S^2$ and $T^2$ that
chirality change in a sense is possible.

We then consider a more physical example, gauge theory
compactified on $T^2$.  If one considers saddle points of the
potential as vacua (nonsupersymmetric and with tachyons), one can
find paths between such vacua which are chirality changing.  This
theory sits in string theory compactified to eight dimensions, so
already one can have chirality change in this extremely simple
example.  It can be made completely transparent by realizing the
gauge theory on Dirichlet branes and doing a T-duality: one has a
simple picture of brane recombination which changes the
intersection number.

In section 4 we discuss aspects of the same problem on the quintic CY.
We exhibit many vector bundles with different values of $c_3(V)$,
which shows that such processes are possible here as well, and discuss
the general problem of finding out which values of $N_{gen}$ can
arise.  We check, using arguments given in the appendix, that there
indeed exists a stable vector bundle with $c_3(V)$ different from that
that of the tangent bundle, which can be used in heterotic
compactification.  Finally, we give an argument (also in the appendix)
that the two term monad construction leads to a finite number of
vector bundles satisfying the anomaly cancellation condition on a
large class of Calabi-Yau manifolds.

In section 5 we discuss a similar brane construction in using
D6-branes in orientifolded type \IIa\ theory, appearing in work of
Cvetic, Shiu and Uranga \cite{CSU} which in fact realizes such a
classical chirality changing transition. We will go a bit further
and prove that in a very similar construction, this transition in
fact proceeds through supersymmetric configurations, and thus
there is no potential barrier.  We suspect this is fairly general
and is true of the construction of \cite{CSU} as well.

Section 6 contains the conclusions and further directions for
study.

\section{Toy models}

We start off by writing down effective Lagrangians which
illustrate the basic point.  In four space-time dimensions, since
the two chiralities of spinor are complex conjugate (or,
equivalently, since one can write Majorana mass terms), chiral
spinors only exist in gauge theory.  The discussion is simplest if
all spinors are taken to be Weyl of one chirality; then a theory
is chiral if the spinors live in a complex representation of the
gauge group.  The same discussion applies to eight space-time
dimensions.

We will write $\CN=1$ supersymmetric effective Lagrangians, being
more relevant to the later discussion.  One could expand in
components and drop fields to get simpler non-supersymmetric
examples.

One can already have complex representations and chiral spinors in
$U(1)$ gauge theories, so these provide the simplest examples. A
simple anomaly free chiral spectrum for $U(1)$ gauge theory is to
take one field $\phi^{(q)}$ with charge $q$, and $q^3$ fields
$\phi^{(-1)}_i$ of charge $-1$. Can we write a single Lagrangian
which allows interpolating between two such spectra with different
$q$ ?

A simple example is the following.  Consider $U(1)\times
U(1)$ gauge theory with chiral multiplets $A$ of charge $(1,0)$,
$B$ of charge $(-1,0)$, $C$ of charge $(0,1)$ and $D$ of charge
$(0,-1)$. We furthermore couple a copy of the chiral spectrum of
our last paragraph with two different values of $q$ to the two
different $U(1)$'s. Finally, we write a superpotential which
forces the constraint
$$
A B + C D = \mu^2
$$
for some scale $\mu$, e.g. by introducing a Lagrange multiplier or
squaring the constraint.

This theory has a moduli space of supersymmetric vacua, and in
particular a vacuum in which $AB=0$, the first $U(1)$ is unbroken
and the first chiral spectrum is realized, and another vacuum in
which $CD=0$, the second $U(1)$ is unbroken and the second chiral
spectrum is realized.

This already varies the chiral spectrum but does not yet give mass
to the chiral multiplets of the ``wrong'' gauge group.  This can
be accomplished by additional superpotential terms; for example in
the $q=2$ theory we can write
$$
W = \sum_i \phi^{(2)}(\phi^{(-1)}_i)^2
$$
which allows lifting almost all of the original chiral matter.
More complicated superpotentials can tie $\phi^{(2)}$ to $ABCD$
above and accomplish the same thing for $q\ne 2$.

The broken $U(1)$ is broken at the scale $\mu$, so if this is
large, a low energy observer will not see the broken $U(1)$.  In
examples arising from Kaluza-Klein compactification, this scale
will be $1/R$, the inverse size of the internal space.

The same mechanism could be used for nonabelian gauge groups as
well, by postulating nonchiral matter analogous to $ABCD$
sufficient to completely break either gauge group.

A more economical mechanism would only require breaking
to a subgroup under which the chiral matter becomes nonchiral. Let
us consider $SO(10)$ grand unified theory containing $N$ spinors
$\psi$ in the $16$. Using the coupling $16\times 16\times 10$, we
can write a coupling to a vector $V_k$ as
\[
W = \sum_{m=1}^N V_k \Gamma^k_{ij}\psi^i_m\psi^j_m
\]
which can lift all the chiral matter at the cost of breaking to
$SO(9)$.  Again, this can be tied to other matter in a way which
restores a different $SO(10)$ elsewhere.

We refrain from going into more detail at this point as there are
very few constraints on this problem from effective field theory.
The only constraint on possible transitions in general $\CN=1$
supersymmetric or non-supersymmetric theories is the rather
trivial one that if the total number of Weyl fermions starts off
even (resp. odd), it must remain even (resp. odd), as fermions can
only be lifted in pairs.  If one requires that some part of the
gauge group remains unbroken or that some chiral multiplets remain
massless along the transition, this could lead to further
constraints, but no specific motivation for such conditions comes
to mind.

\subsection{Kaluza-Klein reduction on $S^1$}

Let us proceed instead by considering specific constructions which
arise from higher dimensional field theory or string theory.

The simplest example is the compactification of $d=5$ gauge theory
with $SU(N)$ vector bundle on $S^1\times\BR^{3,1}$.  Although much
discussed in the literature and surely familiar to most readers,
it is a good warm-up as it already illustrates in what sense the
usual concepts of low energy gauge theory suffice to describe the
situation, and the sense in which the true gauge group contains
infinitely many copies of the low energy gauge group.

Let the $S^1$ have coordinate $x^5$ with periodicity $2\pi$.  A
solution of the Yang-Mills equation on $S^1$ is a flat connection,
whose gauge invariant data is a choice of the holonomy
$W_q=exp\{i\int dx^{5}A_{5}\}$ up to conjugation.  By applying a
gauge transformation in the connected component of the identity
one can bring $A_{5}(x_5)$ to a constant connection
$A_{5}=\vec\theta$, with
$\vec\theta=\mbox{diag}(\theta_1,\theta_2,\ldots,\theta_N)$
and $\sum_i\theta_i=0$.
Permutation of the $\theta_i$ are also gauge transformations (the
Weyl group).  A generic Wilson line will break the gauge group to
$U(1)^{N-1}$.

There are also gauge transformation not continuously connected to
the identity, which can shift any $\theta_i$ by an integer.  Thus
the path $\theta_i=-\theta_j=t$ with $t\in [0,1]$ is a a closed
loop in the moduli space, with unbroken $SU(2)$ symmetry at
$t=0,1$ broken to $U(1)$ at intermediate values.

While in this example the two endpoints are physically equivalent
models, there is a clear sense in which the two $SU(2)$ subgroups
are different embeddings in a larger, infinite dimensional gauge
group, in this case a loop group.  Let us see this by doing the KK
reduction explicitly. The five dimensional gauge connection
decomposes into a spacetime component and an internal space
component, $A_M=(A_\mu, A_5\equiv \phi)$. Choose the internal
gauge connection to be flat and diagonal, as above.
The Fourier expansion of the spacetime components is
\begin{equation}
A_\mu(x^\mu,x_5) =\sum_n A_\mu^{(n)}e^{-inx^5}; \qquad
\phi(x^\mu,x_5) =\sum_n \phi^{(n)}e^{-inx^5} .
\end{equation}
$A_\mu$ and $\phi$ are hermitian, so
$(A_\mu^{(n)})^\dagger=A_\mu^{(-n)}$ (resp. $\phi$). In term of
the matrix element, this is $(a_{ij}^{(n)})^*=a_{ji}^{(-n)}$.  We
expand the parameter $\epsilon(x)$ of local gauge transformations
in the same way.

The four dimensional Lagrangian for the vector bosons is then
\begin{eqnarray*}
\mathcal{L}&=&\int_{S_1} trF_{\mu\nu}^2+ trF_{\mu5}^2, \\
\int_{S_1} trF_{\mu\nu}^2 &=& \sum_n F_{\mu\nu}^{(n)}F_{\mu\nu}^{(-n)}, \\
\int_{S_1} trF_{\mu5}^2 &=& \int_{S_1} tr(\partial_5A_\mu+i[\phi, A_\mu])^2 \\
            &=& \sum_nTr[(-inA_\mu^{(n)}+i[\phi, A_\mu^{(n)}])(inA_\mu^{(-n)}
            +i[\phi, A_\mu^{(-n)}])] \\
            &=& \sum_n \sum_{i,j} [(\theta_i-\theta_j)-n]^2a_{ij}^{(n)}a_{ji}^{(-n)}.\\
\end{eqnarray*}
The $F_{\mu 5}^2$ terms provide mass terms for almost all of the
low energy gauge bosons.  Note that whenever $\theta_i-\theta_j=n$
is an integer, the non-diagonal gauge bosons $A_{ij}^{(n)}$ become
massless, and part of the broken symmetry is restored.  A flow
from $n$ to $n+1$ will however complete the diagonal $U(1)$ gauge
symmetry (for which the gauge bosons are always $a_{ii}^{(0)}$) to
$SU(2)$ using ``different'' off-diagonal gauge bosons
$A_{ij}^{(n)}$ and $A_{ij}^{(n+1)}$.

In the low energy theory, this is just the standard Higgs
phenomenon, as can be seen by considering the gauge
transformation:
$$
\phi(x^5) \rightarrow g(x^5)^{-1} (\p_5 + i \ad \phi(x^5)) g(x^5)
.
$$
Expanded in modes, this is
$$
\delta \phi^{(n)}_{ij} = (n+\theta_i-\theta_j)\epsilon^{(n)}_{ij}
.
$$
If the $\theta_i$ are small, the $\phi^{(n)}$ transform
inhomogeneously under the $n\ne 0$ modes of the gauge symmetry, so
they are the Goldstone bosons for the broken gauge symmetry.  On
the other hand, for general $\theta_i$ one can unbreak
``different'' parts of the gauge group.

Although reduction on $S^1$ is not going to lead to chiral
fermions, one can still see the sense in which massless matter at
the two ends of the loop is ``different'' as well.  The case of
adjoint fermions is the same as above, while fundamental fermions
would decompose into an infinite set of modes $\psi^{(n)}_i$ and
$\tilde\psi^{(n)}_i$ with $U(1)_i$ charge $\pm 1$ and mass
$n-\theta_i$.  Again, any of these modes can appear as massless
matter at a point where non-abelian gauge symmetry is restored.

One can get a more geometric picture of the same result by
considering a gauge theory on Dirichlet branes wrapped on $S^1$
and performing T-duality, leading to Dirichlet branes at points in
the dual $S^1$. The non-abelian gauge bosons arise as winding
strings, and as we move a D-brane around the dual $S^1$,
``different'' winding strings shrink to zero length.  We will use
this picture in other examples below.

We put ``different'' in quotes as the two ends of the paths we are
considering are in fact completely equivalent under large gauge
transformation, a feature which will not be true in the real
examples. In effective field theory, one can get a more
illustrative example by adding additional terms to this effective
Lagrangian which spoil the large gauge symmetry, such as couplings
to additional matter of the sort we postulated earlier.

Naively, one might think of the underlying gauge group as a simple
infinite product of groups ``$\CG=\oprod_{n\in\BZ} G_n$,'' while
to describe a given path one might keep two of these groups (say
$G_0$ and $G_1$ to describe $\theta\rightarrow\theta+1$).  This is
{\it not} a correct interpretation of what we have described.
Rather, the gauge group is a loop group, and the diagonal gauge
bosons always sit in the same Cartan subalgebra.  Thus we have not
described the physics in terms of a product of finite dimensional
Lie groups, but as an infinite dimensional Lie group.

One can certainly write a model which realizes one such path in
terms of two groups $G_0\times G_1$, by postulating additional
bifundamental matter which breaks down to the diagonal subgroup.
However this matter is not present in the underlying KK theory as
we have written it.  One might try to postulate additional matter
in hopes of simplifying the overall description, but it is not
obvious to us that this is more natural than working with the loop
group.

To some extent, the general picture is similar for any
compactification, with the ``true'' gauge group of the effective
gauge theory arising from compactification of Yang-Mills with
gauge group $G$ on an internal space $M$ being the group of maps
from $M$ to $G$.  Unfortunately, it is not easy to get a global
picture of this group for $M$ of dimension greater than 2.  One
furthermore expects larger groups to appear for string scale $M$.
The most concrete version of this we know about appears in the
work of Giveon, Porrati and Rabinovici on $\CN=4$
supergravity-super Yang-Mills theories arising from toroidal
compactification of the heterotic string \cite{GPR}.

\subsection{Connections on $S^2$}

We next discuss $S^2$ and $T^2$.  We will be more interested in
$T^2$, but the story for $S^2$ is simpler and illustrates the
essential points. Although $S^2$ is not Ricci flat, one can
realize gauge theory on this space in string theory, by wrapping
branes on a minimal volume cycle with topology $S^2$.

These spaces can be given complex structures, to become
$\mathbb{CP}^1$ and $\BC/\BZ^2$ respectively, so both cases can
preserve supersymmetry, and both can be discussed in the language
of algebraic geometry.  We will return to this later.  One can
also appeal to the fairly complete mathematical analysis of Atiyah
and Bott \cite{AB}, which finds the general solutions to the
Yang-Mills equations on Riemann surfaces, and discusses how these
sit in the total space of connections. The KK compactification of
$U(1)$ theory coupled with gravity on $S^2$ is discussed in
\cite{SS,RSS}.

One can also do straightforward KK reduction. Let us consider
$S^2$.  On a two dimensional manifold, $U(1)$ bundles are
classified topologically by the first Chern class $c_1(V)$, taking
values in $H^2(M,\BZ)\cong \BZ$. Physically this is a version of
the Dirac quantization condition.  On the other hand, there are no
topological invariants of an $SU(N)$ bundle on a simply connected
two dimensional manifold.

We fix the rotationally symmetric metric on $S^2$ as
$ds^2=d\theta^2+\sin^2\theta d\phi^2$, the corresponding volume
form is $\omega=\frac{1}{4\pi}\sin\theta d\theta d\phi$, with
$\int_{S^2} \omega=1$.

A simple $U(1)$ connection on $S^2$ with $c_1(V)=n\in\BZ$ and
which solves the Yang-Mills equations can be written by
postulating two patches, $B_+\subset S^2$ with $\theta<\pi$ and
$B_-\subset S^2$ with $\theta>0$.  The connections in the two
patches are
\begin{equation} \label{eq:spherecon}
 A_\pm(\theta,\phi)=-\frac{n}{2}d\phi\frac{\cos\theta\pm 1}{\sin\theta} .
\end{equation}
This connection has $F=n\omega$, so $d*F=0$.  We will refer to
this line bundle (and, making a slight abuse of language, this
connection) as $\CO(n)$, for reasons we discuss later.

In fact, all solutions of the Yang-Mills equations on $S^2$ for
any semisimple $G$ can be written as a direct sum of $U(1)$
solutions $\CO(n_i)$, and thus can be specified by an unordered
list of integers $n_i$~\cite{G}.  For $SU(N)$, the only constraint
on these integers is $\sum_i n_i=0$.  We will speak of such an
$SU(N)$ connection as having ``split'' into $U(1)$ connections,
and the unordered list of $n_i$ as the ``splitting type.''

As $SU(N)$ connections, the solutions with $n\ne 0$ are not minima
of the Yang-Mills action (which becomes the four dimensional
effective potential) but are instead saddle points, with unstable
(tachyonic) modes.  This can be seen explicitly, or by appealing
to the results of \cite{AB}, who show that the number of unstable
modes for such a solution is the index of the Yang-Mills
functional at the critical point.  For $\CO(n)\oplus \CO(-n)$ on a
genus $g$ Riemann surface, this is $2(2n-1+g)$.  On $T^2$, it is
easy to see in a T-dual brane description, as we discuss below.

This instability will not prevent us from using these solutions as
simple toy examples of chirality change.  After all, the question
of whether chirality change is possible in effective field theory
does not depend on whether the vacua under discussion are stable
or not.  When we get to the real examples on Calabi-Yau, we will
find that this toy model does properly illustrate the basic
phenomenon.

Rather, the question is whether these different solutions are
topologically connected and lead to different spectra of chiral
fermions in the low energy theory.  Now as $U(1)^{N-1}$
connections, these connections are topologically distinct, and
classified by the homotopy class of the transition function in
$\pi_1(U(1)^{N-1})$ up to Weyl reflection (here $S_N$
permutations).  However, within the space of $SU(N)$ connections,
any of these connections can be deformed into any other; the
transition function is classified by $\pi_1(SU(N)$ which is
trivial.  On the other hand, since all solutions of Yang-Mills
take the form we discussed, all of these deformations must pass
through connections which do not solve the Yang-Mills equations.

An example of a path in the space of $SU(2)$ gauge connections,
corresponding to the deformation from $\mathcal{O}(n)
\oplus\mathcal{O}(-n)$ to $\mathcal{O}(m) \oplus \mathcal{O}(-m)$
as $t$ varies from 0 to 1, can be constructed as follows. Because
the two gauge connections on the two patches can be subject to
arbitrary gauge transformations and their forms are not unique, we
start from the intrinsic description of the vector bundle by gauge
transition function defined at the intersection of the two
hemispheres. Its form is unique up to a constant SU(2)
transformation. This one parameter family of $SU(2)$ valued map on
the equator $S^1$ is
\begin{eqnarray}
 g(t;\phi) &=& \left(\begin{array}{cc}
                           \sin^2\pi t+ \cos^2\pi te^{in\phi}  & -\sin\pi t\cos\pi t(1-e^{in\phi}) \\
                           \sin\pi t\cos\pi t(1-e^{-in\phi}) &  \sin^2\pi t+\cos^2\pi t e^{-in\phi}\\
                           \end{array}\right),
                           \qquad t\in [0,\frac{1}{2}], \\
                  &=& \left(\begin{array}{cc}
                           \sin^2\pi t+ \cos^2\pi te^{im\phi}  & -\sin\pi t\cos\pi t(1-e^{im\phi}) \\
                           \sin\pi t \cos\pi t(1-e^{-im\phi})  &  \sin^2\pi t+\cos^2\pi t e^{-im\phi}\\
                           \end{array}\right),
                           \qquad t\in [\frac{1}{2},1] .
 \end{eqnarray}
This is a single-valued function from $S^1$ to SU(2), and
satisfies all the requirements: at $t=0$ it is of diagonal form
\begin{equation}\label{ONTransFunc}
g(\phi)=\left(\begin{array}{cc}
            e^{in\phi} & 0  \\
            0 & e^{-in\phi} \\
            \end{array}\right),
\end{equation}
which corresponds to the vector bundle $\mathcal{O}(n)\oplus
\mathcal{O}(-n)$, while at $t=1$ for $\mathcal{O}(m)\oplus
\mathcal{O}(-m)$; in the middle $t=\frac{1}{2}$ it is identity.

Let us explain in a bit more detail how this seemingly complicated
expression is obtained. Start from the identity transition
function, a very natural deformation path
\begin{equation}
g'(t;\phi)=\left(\begin{array}{cc}
            \sin\pi t & \cos\pi te^{in\phi}  \\
             -\cos\pi te^{-in\phi} & \sin\pi t  \\
            \end{array}\right).
\end{equation}
At $t=\frac{1}{2}$ this is identity as required. But at $t=0$ this
is not quite the right one: it differs from equation
(\ref{ONTransFunc}) by a constant $SU(2)$ rotation. It can be
easily fixed by multiplying $g'(t;\phi)$ by a coordinate
independent path from identity to this constant matrix
\begin{equation}
g_0(t)=\left(\begin{array}{cc}
             \sin\pi t & -\cos\pi t  \\
            \cos\pi t & \sin\pi t \\
            \end{array}\right).
\end{equation}
This produces the final form of the transition function presented
above.

There are several points worth commenting on.
\begin{itemize}
\item This smooth deformation goes through non-diagonal $SU(2)$
matrices, and therefore through non-diagonal connections. This is
basically required by the single-value condition of the map on the
equator. \item This particular path goes through the trivial
bundle $\mathcal{O}\oplus \mathcal{O}$, as the transition function
at $t=\frac{1}{2}$ is the identity. This fits with the fact that
the trivial bundles is generic in the moduli space of the
$SL(2,C)$ vector bundles. On the other hand, the moduli subspace
which corresponds to nontrivial splittings has nonzero codimension and
is not generic. In particular, any two different splitting types
can be connected continuously through the trivial bundle phase as
we did above. \item It is possible to go directly from
$\mathcal{O}(n)\oplus \mathcal{O}(-n)$ to
$\mathcal{O}(m)\oplus\mathcal{O}(-m)$ by combining two such
deformation together. Specifically, we can construct the
deformation path as
\begin{equation}
g_{n\rightarrow m}(t;\phi)=\begin{pmatrix}
            \sin\frac{\pi}{2} t & \cos\frac{\pi}{2} te^{in\phi}  \cr
             -\cos\frac{\pi}{2} te^{-in\phi} & \sin\frac{\pi}{2} t  \cr
              \end{pmatrix} \cdot
             \begin{pmatrix}
            \sin\frac{\pi}{2} t e^{im\phi}& -\cos\frac{\pi}{2} t \cr
             \cos\frac{\pi}{2} t & \sin\frac{\pi}{2}te^{-m\phi}  \cr
              \end{pmatrix}, t\in [0,1].
\end{equation}
One can easily check that this is a smooth deformation of the
SU(2) transition functions.
\end{itemize}

Next we discuss the gauge connections. The gauge connections on
the two patches are connected by a gauge transformation at the
equator $\theta=\frac{\pi}{2}$ through transition function
$g(t;\phi)$, as $A_+|_{\theta=\frac{\pi}{2}}-g(t;\phi)\cdot
A_-|_{\theta=\frac{\pi}{2}} \cdot g(t;\phi)^{-1} =-\frac{1}{i}
dg(t;\phi) g^{-1}(t,\phi)$. The gauge connections on the two
patches $A_\pm$ can be changed arbitrarily by gauge
transformations. It turns out the "symmetric" choice in which
$A_+|_{\theta=\frac{\pi}{2}}=-A_-|_{\theta=\frac{\pi}{2}}$ does
not give a smooth solution for the connections. But there exists
appropriate choice of gauge such that $A_\pm$ are smooth in
separate patches, which is ensured by the smoothness of
$g^{-1}dg$. This can be achieved by choosing a gauge connection in
one patch, and use $g(t)$ to obtain the connection in another
patch. In particular, one can choose
\begin{eqnarray} \label{eq:connpath}
A_-(t,\theta,\phi) &=& \begin{pmatrix}
                           -\frac{n}{2}\cos\pi t & 0 \\
                           0 & \frac{n}{2}\cos\pi t
                           \end{pmatrix} \frac{\cos\theta
                           +1}{\sin\theta}, t\in [0,
                           \frac{1}{2}]\\
                   &=& \begin{pmatrix}
                           -\frac{m}{2}\cos\pi t & 0 \\
                           0 & \frac{m}{2}\cos\pi t
                           \end{pmatrix} \frac{\cos\theta
                           +1}{\sin\theta}, t\in
                           [\frac{1}{2},1],
\end{eqnarray}
\begin{eqnarray}
A_+(t,\theta,\phi) &=& \frac{n}{2}\cos\pi t K(t;\phi;n)
\frac{\cos\theta -1}{\sin\theta}, t\in [0,
                           \frac{1}{2}]\\
                   &=& \frac{n}{2}\cos\pi t K(t;\phi;m) \frac{\cos\theta
                           -1}{\sin\theta}, t\in [\frac{1}{2},1]
\end{eqnarray}
where $K(t;\pi;n)$ is a $2\times 2$ hermitian matrix smooth in t
and $\phi$, whose explicit form is not illuminating and we will
ignore its detailed form. It smoothly interpolates from
$diag(\frac{n} {2}, -\frac{n}{2})$ at $t=0$ to $0$ at
$t=\frac{1}{2}$ to $diag(\frac{m} {2}, -\frac{m}{2})$ at $t=0$.

%

\subsection{Fermion zero modes on $S^2$ and change of chirality}

The net number of chiral fermion zero modes for a fermion coupling
to the connections we just discussed is determined by the index
theorem,
\begin{equation}\label{eq:twodindex}
\mbox{index}_Q\fs{D}_\Sigma =\frac{1}{2\pi} \int_\Sigma tr_Q F.
\end{equation}
For example, a charge $+1$ fermion coupled to the connection
$\mathcal{O}(n)$ will have $n$ net positive chirality zero modes.

Of course, for a $U(1)$ connection, this number is a topological
invariant.  The corresponding net number of zero modes for an
$SU(n)$ connection is also given by \rfe{twodindex}, which is
zero, and also invariant.

However, if we look at fermion zero modes charged under one of the
$U(1)$ subgroups of $SU(n)$ into which the connections which solve
the Yang-Mills equations split, the net number of these zero modes
can change.  For the path described above, there is a net change
of chiral fermion zero modes, as $t$ varies from 0 to
$\frac{1}{2}$ to 1, the net number of chiral fermions varies from
$n$ to 0 to $n+1$ correspondingly.

Let us see if this effect can lead to change of chirality in the
low energy theory.  Let us take $SU(2)$ for definiteness.  Each of
the solutions $\CO(n)\oplus \CO(-n)$ has (naively) the same
unbroken gauge symmetry $U(1)$ with the same embedding, which we
can describe in terms of the infinitesimal gauge transformation
\begin{equation}\label{eq:subgroup}
\delta A_\mu = \left(\begin{matrix} 1&0\cr 0&-1
\end{matrix}\right)
 \partial_\mu \epsilon .
\end{equation}

If we consider fermions in the fundamental representation of
$SU(2)$, the two components couple to $\CO(n)$ and $\CO(-n)$ and
respectively have charges $+1$ and $-1$ under the low energy
$U(1)$ gauge group. A fermion in the adjoint of $SU(2)$ will
decompose into one neutral fermion, and two off-diagonal
components coupling to $\CO(2n)$ and $\CO(-2n)$ with respective
low energy $U(1)$ charges $+2$ and $-2$ (in the same units).

The discussion for the two cases is qualitatively the same, with a
correlation between the internal and low energy $U(1)$ charges,
which is preserved under complex conjugation.  Because of this,
the chiral fermions predicted by (\ref{eq:twodindex}), which might
naively be thought to cancel between $\CO(n)$ and $\CO(-n)$, in
fact do not cancel from the point of view of low energy chirality.

Consider the reduction of $5+1$ gauge theory for definiteness, and
decompose $SO(1,5)$ into $SO(1,3)\times SO(2)$.  A chiral spinor
in $d=5+1$ (say a $4$-component complex Weyl spinor) reduces to
$2_1 \oplus \bar 2_{-1}$.  Because the $d=6$ spinor is complex,
the two terms in this decomposition are not complex conjugate but
instead are independent $d=4$ spinors, with correlated chirality
in $d=4$ and $d=2$.

We can now apply the index theorem as above to conclude that a
$d=6$ Weyl spinor in the fundamental leads to $n$ left-handed
$d=4$ spinors with $U(1)$ charge $+1$, and $n$ right-handed $d=4$
spinors with $U(1)$ charge $-1$.  Complex conjugating the latter,
we obtain a chiral spectrum of $2n$ spinors of charge $+1$.
(Because the original fermion was complex, these are not CPT
conjugate.) Similarly, the adjoint representation leads to $2n$
left handed spinors of charge $+2$.

The case of reduction from ten to eight dimensions is similar.
Starting with a Majorana-Weyl ten dimensional spinor, we make the
decomposition into $SO(7,1)\times SO(2)$ spinors $16=8_1 \oplus
8'_{-1}$, which is different from $d=6$ only in that the original
spinor is real, so the two terms in the decomposition are complex
conjugate to each other.  The rest of the argument is the same,
leading to a chiral spectrum in $d=7+1$ with multiplicities half
of what we found for $d=3+1$.

These results seemed rather counter-intuitive to us at first so
let us discuss some of the objections which may come to mind.
First, the two chiralities of spinor in $d=5+1$ and $9+1$ are not
complex conjugate, so there is no {\it a priori} argument that one
must start with a complex representation of the gauge group to get
chirality.  This allowed us to start with $SU(2)$ and even the
adjoint representation and obtain a chiral spectrum in lower
dimensions.

One can also notice that the resulting four dimensional spectra
are anomalous.  This is no contradiction because the six and ten
dimensional theories were also anomalous.  Since our arguments are
classical, this need not concern us.  Interesting examples will of
course cancel the higher dimensional anomalies, as will happen in
string theory, and then the resulting four dimensional spectrum
will be non-anomalous.

Finally, for very skeptical readers, there are brane arguments
later which may make this result more believable.

\subsubsection{What is going on}

One can use the result we just obtained to get a four dimensional
effective theory in which by varying the vacuum expectation value
of a field, the parameter $t$ of our family of connections, one
can interpolate between theories with different numbers of  $U(1)$
charged chiral multiplets.

According to our previous discussion, this does not contradict the
standard wisdom, as long as the initial variation of the field
breaks the original $U(1)$ gauge group, and a different $U(1)$ is
restored at the endpoint.

What is confusing, is that it appears that both unbroken $U(1)$'s
at the two ends of the path have the same embedding \rfe{subgroup}
in the $SU(2)$, and thus should correspond to the {\it same}
$U(1)$ gauge group in the effective theory.  More precisely, a
$U(1)$ gauge boson in the low energy theory arises from a
covariantly constant scalar
$$
D_i \Phi = \p_i\Phi + [A_i,\Phi] = 0
$$
in the higher dimensional theory as
$$
A_\mu(x,\theta,\phi) = A_\mu(x) \Phi(\theta,\phi) .
$$
Since the connection $\CO(n)\oplus \CO(-n)$ can be written in
terms of \rfe{spherecon} as
$$
A_i =
 \left(\begin{matrix} A_{\pm}(\theta,\phi)
&0\cr 0&-A_{\pm}(\theta,\phi) \end{matrix}\right) ,
$$
each of these connections admits a covariantly constant scalar
\begin{equation}\label{conscalar}
\Phi(\theta,\phi) =
 \left(\begin{matrix} 1&0\cr 0&-1 \end{matrix}\right)
\end{equation}
which are ``the same,'' leading to a paradox.

The subtlety which resolves this is that to properly compare
sections of the two bundles, we must work with a presentation in
which the transition functions are the same.  However, the section
(\ref{conscalar}) is defined using transition functions
(\ref{ONTransFunc}) which are different for each $n$.

Since all of these $SU(2)$ bundles are topologically trivial,
there is no need to work with patches and transition functions to
describe their sections.  Concretely, if we can extend the gauge
transformation on the equator to (say) the southern patch, we can
use it to convert the sections of the adjoint bundle to functions
on the sphere, and compare these.

Explicitly, for the bundle $\mathcal{O}(n)\oplus\mathcal{O}(-n)$,
we extend the transition function to the southern patch as
\[
g_+(\theta, \phi) =\begin{pmatrix} \cos^2\theta+\sin^2\theta
e^{in\phi}
& -\frac{1}{2}\sin 2\theta(1-e^{in\phi})\\
\frac{1}{2}\sin 2\theta(1-e^{-in\phi}) & \cos^2\theta+\sin^2\theta
e^{-in\phi}  \end{pmatrix}.
\]
Then the adjoint scalar zero mode $\Phi$, as a function on the
sphere, is
\begin{eqnarray*}
\Phi(n;\phi) &=& \begin{pmatrix} 1 & 0 \\ 0 & -1 \end{pmatrix}, \qquad \theta\in [0,\frac{\pi}{2}], \\
     &=& \begin{pmatrix}
          1-\sin^22\theta(1-\cos n\phi) & \sin 2\theta(1-e^{in\phi})(\cos^2\theta+\sin^2\theta e^{in\phi}) \\
          \sin 2\theta(1-e^{in\phi})(\cos^2\theta+\sin^2\theta e^{in\phi}) &  -1+\sin^22\theta(1-\cos n\phi) \end{pmatrix},  \theta\in [0,\frac{\pi}{2}].
\end{eqnarray*}
Obviously, two such functions $\Phi(n;\phi)$ and $\Phi(m;\phi)$ on
$S^2$, which correspond to zero modes of
$\mathcal{O}(n)\oplus\mathcal{O}(-n)$ and
$\mathcal{O}(m)\oplus\mathcal{O}(-m)$ respectively, are different.
This shows that the zero modes at the two ends of the deformation
are different, and so are the $U(1)$ symmetry groups in the
effective theory.

One might worry that this argument depends on the choices of
extension. To see that it is independent of such choices, observe
that the function $\Phi$ at the southern patch is
$g_+\begin{pmatrix} 1 & 0
\\ 0 & -1
\end{pmatrix} g_+^{-1}$. If $\Phi(n)=\Phi(m)$, then on the southern patch one has
\[g_+(m)\begin{pmatrix} 1 & 0
\\ 0 & -1
\end{pmatrix} g_+(m)^{-1}=g_+(n)\begin{pmatrix} 1 & 0
\\ 0 & -1
\end{pmatrix} g_+(n)^{-1}, \]
which requires $g_+(n)g_+^{-1}(m)$ commutes with $\begin{pmatrix}
1 & 0 \\ 0 & -1 \end{pmatrix}$. This is possible only if it is a
diagonal matrix. Now $g_+(n)g_+^{-1}(m)$ is a $SU(2)$ matrix
interpolating from $\begin{pmatrix} e^{i(n-m)\phi} & 0 \\
0 & e^{-i(n-m)\phi}
\end{pmatrix}$ at $\theta=\frac{\pi}{2}$ to the identity matrix at
$\theta=\pi$, as $\theta$ varies from $\pi/2$ to $\pi$. A diagonal
$SU(2)$ matrix is isomorphic to $U(1)$ which is topologically
$S^1$. Since it is impossible to interpolate between two maps
$S^1\rightarrow S^1$ with different winding number,
$g_+(n;\theta)g_+^{-1}(m;\theta)$ must have non-diagonal elements
at certain $\theta$ values. Hence $\Phi_n\ne\Phi_m$ follows
immediately.

Thus we have seen in detail how a continuous path between two
backgrounds solving the Yang-Mills equations on $S^2$ can
interpolate between low energy theories with different net numbers
of $U(1)$ charged chiral fermions, again because different $U(1)$
subgroups of the underlying infinite dimensional gauge group are
unbroken. One can proceed using KK reduction to develop the
effective field theory description, which is most simply done by
describing the vector bundles without using transition functions,
to obtain an infinite component field theory similar to what we
saw for $S^1$.

\subsection{The torus $T^2$ and the brane picture}

The story for $T^2$ is quite similar. One again obtains a large
class of solutions to Yang-Mills as direct sums of line bundles.
Now, such a solution is not classified only by the first Chern
class, but also has two real parameters (for a flat connection
this is of course the Wilson line). The solutions corresponding to
each splitting type then form a continuous moduli space, as
opposed to the case of $S^2$ where the solution corresponding to
each splitting type is merely a point. Apart from this difference,
all other qualitative features are the same. The moduli spaces
corresponding to different splitting types are disconnected as
Yang-Mills solutions, but they can be connected through a path via
gauge connections that are non-solutions. One again finds that
direct sums of $U(1)$ connections with non-zero field strength are
(unstable) solutions of Yang-Mills with chiral fermion zero modes,
whose number is given by the same formula (\ref{eq:twodindex}).
Chirality change is possible because the unbroken subgroup,
associated to a covariantly constant scalar $D_i \Phi = 0$, is
different for the different saddle points (they can be written
explicitly using theta functions).

The $T^2$ picture can be made much more intuitive by realizing the
gauge theory on branes and taking advantage of T-duality. For
definiteness let us consider type \I\ string theory with $32$
D9-branes compactified on $T^2$. The maximal gauge group is
$SO(32)$.

The simplest picture is obtained by T-dualizing a single
coordinate, to obtain ``type \I' theory'' with $32$ D8-branes and
two O8-planes. The T-duality exchange the K\"{a}hler moduli
parameter $\tau$ and the complex structure parameter $\rho$, and
in general turns a torus with nondiagonal metric into a torus with
constant background $B$ field. To avoid complication due to
constant $B$ flux, we restrict to orthogonal torus (so the metric
is diagonal and $\rho$ is pure imaginary) which is mapped into
itself after T-duality.

Let us consider a simple example which embeds a $U(1)$
connection with nontrivial field strength in the $SO(32)$ gauge
group. Assume the field strength takes the form
$$
F=\left(\begin{matrix}
0& k& 0& 0& 0& 0& \ldots\\
-k&0& 0& 0& 0& 0& \ldots\\
0& 0& 0& k& 0& 0& \ldots\\
0& 0&-k& 0& 0& 0& \ldots\\
\hdots& \hdots&\hdots& \hdots&\hdots& \ddots
\end{matrix}\right)
\ dx\wedge dy .
$$
The low
energy effective theory is a chiral gauge theory with gauge group
$SU(16)\times U(1)$, and $k$ copies of chiral fermions in the
fundamental representation of $SU16)$, as follows from the index
theorem.

To relate this to our previous discussion, it is simplest to think of
$SO(32)$ as embedded in $SU(32)$, postulate the $SU(32)$ connection
$\CO(k)^{16}\oplus\CO(-k)^{16}$, and then impose a reality condition
on the fields.  This parallels the usual construction of type \I\
string using the $\Omega$ projection.  There is much more to say about
this, which we discuss elsewhere.

After T-dualizing in the $y$ direction, the torus is an orbifold
$S^1/Z_2\times S^1$. We work in the covering space $T^2$. Let
$Z_2: y\rightarrow -y$, any brane configuration on orbifold is
lifted to $Z_2$-invariant configuration. The gauge connection in
the $y$ direction become the $y$ position of the branes in the
dual picture. Thus we have 16 branes described by $y=kx$ and 16
brane images described by $y=-kx$, or in terms of the A- and
B-cycles, they wraps $(1,k)$ and $(1,-k)$ respectively. There are
$2k$ intersection points between the two sets of branes,
corresponding to k CPT pairs of chiral fermions in the $16$ and
$\bar{16}$ representation of $SU(16)$.

\begin{figure}
\begin{center}
\centering \epsfysize=3.5cm \leavevmode
\epsfbox{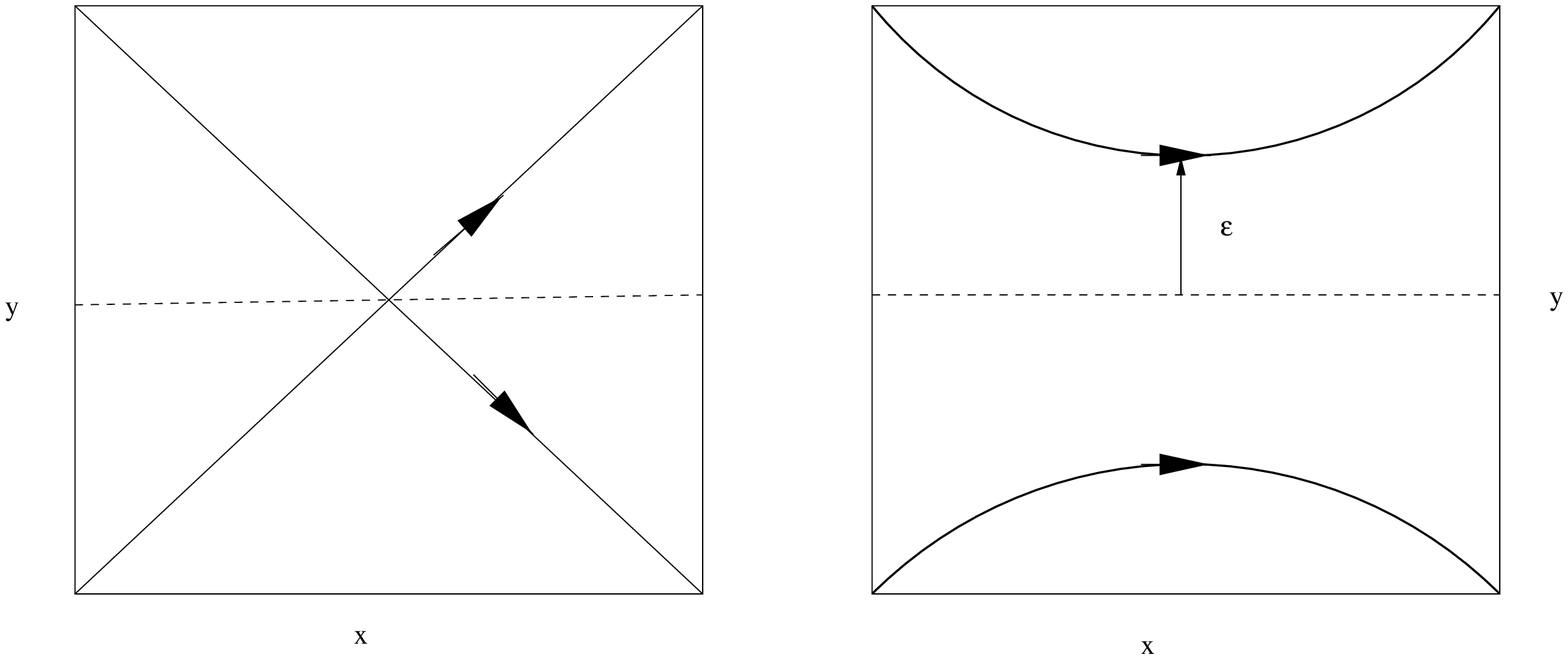}
\end{center}
\caption[]{\small Two intersecting D8 branes deform to a
configuration without intersection, which realizes the change of
chirality change.} \label{braneA}
\end{figure}

A geometric picture of smooth change of chirality from $k=1$ to
$k=0$ is shown in fig.\ref{braneA}. The two sets of branes wrap
the one cycle $[a]+[b]$ and $[a]-[b]$ respectively. As one deforms
the branes, the chiral fermions living at the two points of
intersection of the $k=1$ branes pair up and gain a mass
proportional to the distance between the two sets of branes.  The
intermediate configurations are not minimal surfaces, and the
corresponding gauge connections do not satisfy the Yang-Mills
equation. Furthermore, they will be irreducible $SU(2)$
connections, and the gauge group in the effective theory is broken
to $SU(16)$. Finally when one reaches a minimal configuration
again (all the branes coincide with the O8-branes at one end of
$S^1/Z_2$) the gauge group is restored to $SO(32)$.

Again sice the change of chirality (in the case we just described
it is actually disappearance of chiral fermions) is accompanied
with the change of gauge group, it can be understood as the Higgs
phenomena.

For general $k$ the picture is essentially the same. We see that there
are an infinite number of brane configurations, parameterized by $k\in
Z^+\cup\{0\}$ and each carries $2k$ chiral fermions, each connected by
brane deformation.  There are even more configurations, constructed by
wrapping branes about the cycles $n[a]+k[b]$.  For $g.c.d(n,k)=1$,
these are T-dual to irreducible connectionis with rank greater than
one \cite{GSToron}.

In this example, the infinite number of possibilities is clearly
related to the fact that we are considering unstable configurations.
The only consistent stable configurations are those in which all
branes are parallel, which prevents the type of cancellations in
the tadpole condition we are seeing here, and it is easy to see
that these are finite in number.  One hopes that this finiteness
is true in higher dimensions as well.

\subsection{Holomorphic bundles, splitting type, and Yang-Mills connections}

To get a similar understanding in Calabi-Yau compactification, we
will need better mathematical technology.  Most of our
understanding in this case is for the special case of
supersymmetric compactification of string theory, for which one
can appeal to algebraic geometry and the theory of holomorphic
bundles.

Within the set of all vector bundles on a complex manifold, one
can distinguish the holomorphic bundles, those for which (in some
basis) the transition functions are holomorphic.  The necessary
condition for this is $F^{(0,2)}=0$, so in one complex dimension
any vector bundle can be treated this way.

This is a great mathematical advantage as the problem of finding
holomorphic bundles does not require solving differential
equations; it is purely algebraic.  One can then go on to solve
the Yang-Mills equations $D*F^{(1,1)}=0$.  To be a bit more
precise, for holomorphic bundles the natural concept of gauge
equivalence is more general; one allows general gauge
transformations $g(z)\in GL(N,\BC)$ (or more generally a
non-compact group obtained from the complexification of the
original Lie algebra), and the statement is that within this
larger gauge orbit we can look for specific unitary connections
which solve the Yang-Mills equations, here equivalent to
$F^{(1,1)} = c\omega$.  When one exists, it is unique up to
standard (unitary) gauge equivalence; the necessary and sufficient
condition for it to exist is $\mu$-stability, as discussed in
great detail in \cite{OSS,PO}.

A rank 1 holomorphic bundle is usually called a line bundle. The
line bundles on $\CP^1$ are classified by the integer $n=c_1(V)$
(the degree), and can be written in terms of two patches and a
transition function $g=z^n$.  The usual notation for such a line
bundle is $\CO(n)$.  One can equivalently say that
$\CO_{\CP^d}(n)$ for $n\ge 0$ has as holomorphic sections
functions of degree $n$ in the homogeneous coordinates on $\CP^d$.
The bundles with $n<0$ can be defined as the duals of these, or as
bundles whose sections must have poles of total order $-n$.

As for $GL(N)$ bundles on $\CP^1$, these are classified by
Grothendieck's theorem: a holomorphic vector bundle $V$ can always
be written in as a direct sums of line bundles,
$$
V = \oplus_i \CO(n_i)
$$
where the $n_i$ are uniquely determined (again as an unordered list).%
One sometimes says that the bundle ``splits,'' and this
decomposition is its ``splitting type.'' Note that this is a finer
classification than by topological type, and in fact is the same
as our classification of solutions of the Yang-Mills equations,
for reasons we will explain shortly. The theorem also classifies
$SL(N)$ bundles: these are just the subset with $\sum_i n_i=0$.

As we said, on $S^2$ (or any two dimensional compact manifold) all
$SL(N)$ bundles are continuously connected and are all equivalent
to holomorphic bundles.  Thus it must be possible to make
continuous transitions between elements of this discrete set of
bundles.  This seeming paradox is not hard to understand. Consider
a family of holomorphic bundles obtained by varying a complex
parameter $\alpha$, defined by the transition
functions \cite{PS,DK}
\begin{equation}
g(z) = \left(\begin{matrix} z&\alpha\cr 0& z^{-1} \end{matrix}
\right).
\end{equation}
For $\alpha=0$ this is $\CO(1)\oplus \CO(-1)$.  On the other hand,
for any $\alpha\ne 0$ it turns out that this is equivalent as a
holomorphic bundle to $\CO \oplus \CO$. This can be seen from the
decomposition
$$
\begin{pmatrix}  z&\alpha\cr 0& z^{-1} \end{pmatrix} =
\begin{pmatrix}  0&\alpha\cr \alpha^{-1}& z^{-1} \end{pmatrix} \cdot
\begin{pmatrix}  -1&0\cr \alpha^{-1}z& 1 \end{pmatrix}.
$$
The two matrices can be extended \emph{holomorphically} to the
unit disk centered around $z=0$ and $z=\infty$ respectively. The
transition function on the equator is trivial up to gauge
transformation on the two open patches, thus corresponds to
trivial splitting types of the line bundles.

This phenomenon is generic and is called ``jumping''. A
simple example of a compact family of bundles which exhibits it
can be constructed by considering a particular non-uniform rank-2
vector bundle $E$ on $P^2$\cite{OSS}. Since $P^1$ can be embedded
as a line in $P^2$, in a family parameterized by the Grassmannian
manifold $G(P^2,P^1)$, the pullback of $E$ to $P^1$ of embedded
yields a family of rank 2 vector bundles on $P^1$, whose spitting
type over the moduli space $G(P^2,P^1)$ demonstrates the jump of
degree. More explicitly, blow up $m$ points $x_1,\cdots,x_m$ on
$P^2$ to get $Y$ with projection $\sigma: Y\rightarrow P^2$. On
$Y$, construct a rank two vector bundle $E'$ via extension
\begin{equation}
 0\rightarrow [C]\rightarrow
E' \rightarrow [-C] \rightarrow 0.
\end{equation}
where $C$ is the exceptional divisors from the blown up. It can be
proved that $E'$ actually is a lift of a rank two vector bundle
$E$ over $P^2$, $E'= \sigma^* E$. $E'$ restricted to the
exceptional divisor $C_i$ are of the form
\begin{equation}
  0\rightarrow \mathcal{O}_{C_i}(-1)\rightarrow
\mathcal{O}_{C_i} \oplus \mathcal{O}_{C_i} \rightarrow
\mathcal{O}_{C_i}(1) \rightarrow 0.
\end{equation}
 Because these exceptional curves become points in $P^2$, when the line
 $P^1$ pass through any point $x_i$, we expect a splitting of E as
 $\mathcal{O} \oplus \mathcal{O} \rightarrow \mathcal{O}(1)\oplus \mathcal{O}(-1)$.
 This is indeed what happens. The general theorem states that the restriction
 of the rank 2 vector bundle E to any line $L$, on which exactly k
 points of the set
 $\{x_1, \cdots x_m\}$ lie, splits in the form
\begin{equation}
E|_L=\mathcal{O}_L(k)\oplus \mathcal{O}_L(-k).
\end{equation}
The generic splitting type of this bundle is
$\mathcal{O}_{P^1}\oplus \mathcal{O}_{P^1}$.

Any rank one line bundle on $\CP^1$ is $\mu$-stable and thus
admits connections solving the Yang-Mills equations,
\rfe{spherecon}. Thus, all of the considerations of the previous
sections using explicit gauge connections can be translated to
this holomorphic context.  We will see some of this in a
Calabi-Yau example below.

To avoid confusion, one should keep in mind that despite the fact
that the holomorphic bundles of different splitting type are
connected by an infinitesimal variation of parameter $\alpha$, the
corresponding solutions of the Yang-Mills equations are separated
by a finite distance.  We saw this explicitly in the previous,
non-algebraic geometry analysis.  In general, a family of
solutions of hermitian Yang-Mills, only arises from a family of
holomorphic bundles.

\section{Chirality change on compact Calabi-Yau in heterotic string theory}

Having understood chirality change in two dimensions, we proceed
to six-dimensional compact Calabi-Yau manifold, keeping in mind
the idea that isomorphic gauge groups can have different
embeddings in the same yet higher dimensional gauge group.

By now the reader is probably willing to grant the argument we
gave in the introduction, that since $c_3(V)$ is not a topological
invariant of $E_8\times E_8$ or $Spin(32)/\BZ_2$ bundles, there is
no topological barrier to changing the number of generations. We
still need to check that in fact there are compactifications on
the same CY with different numbers of generations which could be
connected in this way.

To find such supersymmetric compactifications, we just need to
find examples of semi-stable holomorphic bundles on the same CY
whose Chern classes differ only in $c_3$.  This is more or less
already known from the study of $(0,2)$ models and is in any case
easy to show using the techniques commonly used there, as we will
do.

We will then proceed to look for something stronger -- a path
between two vacua with different numbers of generations completely
through supersymmetric configurations.

\subsection{Chirality from reducible connections}

Consider heterotic string theory with gauge group $E_8\times E_8$
compactified on a simply connected compact Calabi-Yau manifold
$M$, and assume the gauge connection on Calabi-Yau is embedded in
the first $E_8$ group.
$\CN=1$ supersymmetry in spacetime effective theory then requires
the following:

(1) The Bianchi  identity for the 3-form field strength $H$ of the
2-form $B$ field $dH=\tr R\wedge R-\tr F\wedge F$ requires that
vector bundle $W$ and the tangent bundle
of the Calabi-Yau manifold $TM$ to have same second Chern classes.
More generally, this
equation also receives a five-brane contribution, but we will not
use this.

(2) The gauge connection $A$ on vector bundle $W$ should satisfy
the Hermitian-Yang-Mills equation, $F_{\bar{a}\bar{b}}=F_{ab}=0$
and $g^{a\bar{b}}F_{a\bar{b}}=0$. By the Donaldson-Uhlenbeck-Yau
theorems, such a solution exists if and only if the vector bundle
$W$ carrying representation of $A$ is a direct sum of stable
bundles with slope zero granting $c_1(W)=0$ (for more on this see the
appendix).

Note that the topological type of $W$ as an $E_8$ bundle is
completely fixed by $c_2(W)$, which we fixed in (1). This is
because the homotopy groups $\pi_i(E_8)$ vanish for all $i\ne 3$
and $i\le 6$ (in fact, for all $i\le 14$).  Thus all such compactifications
are connected, in general through non-supersymmetric configurations.

Of course, compactification with an $E_8$ connection leaves no chiral
fermions (and no unbroken gauge group) perturbatively.
To get a non-trivial low energy gauge group
$G$, we choose a subgroup $H\subset E_8$ and a bundle $V$ with
structure group $H$; $G$ is then the commutant of $H$ in $E_8$.
The moduli space of such $V$ is a subset of the total moduli space
of bundles $W$, and in the neighborhood of $V$ one can think of
this condition as simply setting the $G$-charged fields to zero.

$W$ can then be decomposed into a direct sum of vector bundles
carrying representations of $G\times H$,
and one can then use the
index theorem to find the chiral spectrum.

The index theorem for a Dirac
operator taking values in a representation $Q$ of the gauge
connections on a six-dimensional manifold X is
\begin{equation}
\mbox{index}_Q \fs{D}_X =\frac{1}{48(2\pi)^2}\int_X[tr_QF\wedge F\wedge
F -\frac{1}{8}tr_QF\wedge trR\wedge R].
\end{equation}
For a vector bundle with $c_1=0$, as will be the case for $H=SU(k)$,
and the fundamental representation (as for GUT matter in the $27$,
$16$, or $10$ with $k=3,4,5$),
this reduces to the third Chern class of the gauge bundle,
and we find $N_{gen}=c_3(V)/2$.

We repeat this well-known material mostly to make the point that there
is a direct analogy between $c_3(V)$ in this construction, and
a quantity such as $k$ in $O(k)\oplus O(-k)$ in our toy models.  It is
not a topological invariant of the $E_8$ bundle, but rather is
directly analogous to the idea of ``splitting type'' we discussed
earlier.

\subsection{Finiteness of the number of possible $c_3$'s}

A very central question in superstring compactification is whether
there are finitely or infinitely many possible compactifications which
might be candidates to describe our universe.  At present even very
basic questions of this type remain unanswered; for example whether
or not the number of Calabi-Yau threefolds is finite.

A question of this type for which the answer is actually known is
whether the $SU(k)$ bundles which satisfy the above requirements
for heterotic string compactification, come with a finite or
infinite number of distinct values of $c_3(V)$. In fact, there is
a finite list of possible values.\footnote{ M.R.D. thanks F.
Bogomolov and A. Langer for explanations on the content of this
subsection.} This follows for a theorem of Maruyama
\cite{Maruyama}, and more recent work of Langer\cite{Langer},
which show that, having fixed the rank, $c_1$ and $c_2$, there are
a finite number of possible values of $c_3$ taken by semistable
bundles.

An intuitive explanation of the result is the following.  One
considers a hypersurface $\Sigma$ in the CY3, and considers the
restriction of $V$ to $\Sigma$.  One can show that, for sufficiently
high degree $\Sigma$, distinct bundles $V_1$ and $V_2$ must restrict
to distinct bundles on $\Sigma$ (this follows from considering the
exact sequence governing the restriction of $\Hom(V_i,V_j)$ to
$\Sigma$), so the restriction actually determines the original bundle
on $V$. Next, note that on restriction, the bundle on $\Sigma$ can
inherit all of the information in the rank, $c_1$ and $c_2$, but loses
all knowledge of $c_3$.  Finally, there is a general result called
``algebraicity of the family,'' which states that a family of bundles
for which all Chern classes are specified is algebraic, roughly
meaning that its moduli space can be defined by a finite system of
equations and thus consists of a finite set of branches.  Thus, the
combined moduli space of all bundles with all values of $c_3$ consists
of a finite set of branches, and thus there must be a finite number of
possible values of $c_3$.

Unfortunately, there is no effective version of this theorem, which
gives actual bounds on the values of $c_3$ which can be attained.

\subsection{Explicit search on the quintic}

We first note that we can construct sheaves on the ambient
projective space, and restrict these to the quintic. This does not
give us all bundles; for a start it imposes conditions on the Chern classes.
For example, on the quintic, we will only get $c_3(V)$ divisible by five.
Furthermore, we search only
the class of vector bundles produced by the two term monad
construction on weighted projective space. There is a more general
three term monad which is the general class of bundle produced by the linear
sigma model.  More general complexes are possible as well; on general
grounds one expects to need at least four terms to get all bundles.

We start a bit more generally by considering intersections of hypersurfaces in
weighted projective space.  If we do not need to resolve singularities,
the resulting even cohomology space has $\mbox{dim} H^{1,1}(M)=\mbox{dim}
H^{2,2}(M)=1$, so the anomaly cancellation condition
imposes a minimal constraint.  The
bundles we are seeking satisfy conditions
\begin{equation}
c_1(V)=0,\qquad  c_2(V)=c_2(TM),\qquad  c_3(V) \ \mbox{arbitrary}.
\end{equation}
The monad construction of vector bundles is studied in~\cite{DB,
K, CLSMonad, H, LO}.

Weighted projective space $CP^{N+3}_w$ has homogeneous coordinates
$(X_1, X_2, ..., X_{N+3})$ with weights $w=(w_0, w_1, ...,
w_{N+3})$ under a $U(1)$ action. A 3-fold is defined as zero locus
of N polynomials $P_1(X)$, $P_2(X)$, $\cdots$, $P_N(X)$ with
degrees $p=(p_1, p_2, ..., p_N)$. The Calabi-Yau condition
requires the first Chern class to be zero,
\begin{equation}
c_1(TX)=(\sum_{i=0}^{N+3}w_i-\sum_{j=1}^N p_j)J,
\end{equation}
where $J=c_1(\mathcal{O}(1))$ is the first Chern class of the
hyperplane bundle on the projective space. The Calabi-Yau
condition requires $c_1(TX)=0$, so
\begin{equation}
\sum_{i=0}^{N+3}w_i=\sum_{j=1}^N p_j.
\end{equation}
Using this condition, the second and the third Chern class can be
expressed as
\begin{eqnarray} c_2(TX) &=& \frac{1}{2}(\sum_{j=1}^N
p_j^2-\sum_{i=0}^{N+3}w_i^2)J^2,\\
c_3(TX) &=& \frac{1}{3}(\sum_{i=0}^{N+3}w_i^3-\sum_{j=1}^N
p_j^3)J^3.
\end{eqnarray}

The monad construction of rank-n vector bundle V is defined by a
short exact sequence
\begin{equation}\label{eq:MONAD}
 0\rightarrow V \rightarrow \oplus_{i=1}^{n+M}\mathcal{O}(n_i)
 \stackrel{\phi=(Q^i_j(X))}{\longrightarrow}
 \oplus_{j=1}^M\mathcal{O}(m_j)\rightarrow 0.
\end{equation}
where the map $\phi$ is defined by a $M\times(M+n)$ matrix of polynomials
$Q^i_j(X_1, X_2, \cdots, X_{N+3}) (i=1, \cdots, M+n; j=1, \cdots,
M)$ with degree $deg(Q^i_j)=m_j-n_i$. Obviously,
\[m_j>n_i>0, \qquad \forall i,j,\]
otherwise the corresponding polynomial is zero. Notice that the
degree is in terms of $\mathcal{O}(1)$, the hypersurface on
$CP_W^{N+3}$.

The first Chern class for such a bundle is
\begin{equation}
c_1(V)=(\sum_{i=0}^{n+M}n_i-\sum_{j=1}^M m_j)J.
\end{equation}
One of the necessary condition for semi-stability is $c_1(V)=0$,
which requires
\begin{equation}
\sum_{i=0}^{n+M}n_i=\sum_{j=1}^M m_j.
\end{equation}
Using this condition, the second and the third Chern classes of V
can be expressed as
\begin{equation}
 c_2(V)= \frac{1}{2}(\sum_{j=1}^M m_j^2-\sum_{i=0}^{n+M}n_i^2)J^2, \qquad
  c_3= \frac{1}{3}(\sum_{i=0}^{n+M}n_i^3-\sum_{j=1}^M m_j^3)J^3.
\end{equation}

To conclude, on a Calabi-Yau space $M$ defined as intersection of
hypersurfaces in the weighted projective space specified by
weights $w_a(a=1,\cdots, N+3)$ and degree of polynomials
$p_b(b=1,\cdots, N)$, define a constant
\begin{equation} \label{QValue}
B=\sum_{j=1}^N p_j^2-\sum_{i=0}^{N+3}w_i^2,
\end{equation}
such that the second Chern class of the tangent bundle is
$c_2(TM)=\frac{1}{2}B\cdot J^2$. The rank-n vector bundle V is
specified by two sets of integers $(n_i;m_j)$ constrained by
\begin{equation}
\sum_{i=0}^{n+M}n_i=\sum_{j=1}^M m_j, \qquad
\sum_{i=0}^{n+M}n_i^2+B =\sum_{j=1}^M m_j^2, \qquad m_j>n_i>0,
\qquad \forall i,j.
\end{equation}
After solving these constraints one can calculate the
corresponding $c_3(V)$.

First, there exists at least one solution to such sets of
constraints, namely, the tangent bundle of the Calabi-Yau manifold
itself. The next qualitative question is whether the number of
solutions to these constraints is finite or infinite. In fact,
as we argue in appendix B, the
number of solutions to these constraints is always finite.

We performed an exhaustive search for the case of the
quintic hypersurface for data solving the above constraint equations.
and rank up to $8$.
The result is shown in Table\ref{QuinticBundle}.
The number of generations is given by
$N_{gen}=\frac{1}{2}C_3(V)$.

\begin{table} \footnotesize
\begin{center}
\begin{tabular}{|c|c|c|}
  \hline\hline
  rank (n) & $(n_i, m_j)$ & $N_{gen}$ \\
    \hline\hline
  3 & (22222222;33334) & 90 \\
    \hline
  3 & (122222;344) & 95 \\
  \hline
  3 & (112233;444) & 100 \\
    \hline
  3 & (11222;35) & 105 \\
    \hline
  3 & (11133;45) & 110 \\
  \hline
  4 & (1122222222;333333) & 80 \\
    \hline
  4 & (11122222;3334) & 85\\
    \hline
  4 & (111122;44) & 90\\
    \hline
  4 & (11111;5) & 100 \\
    \hline
  5 & (1111122222;33333) & 75 \\
    \hline
  5 & (11111122;334) & 80 \\
    \hline
  6 & (1111111122;3333) & 70 \\
    \hline
  6 & (111111111;234) & 75 \\
    \hline
  7 & (11111111111;2333) & 65 \\
    \hline
  7 & (111111111111;22224) & 70 \\
    \hline
  8 & (11111111111111;222233) & 60 \\
    \hline
\end{tabular}
\end{center}
\caption{\small List of holomorphic vector bundles on quintic
Calabi-Yau manifold with $c_1=0, c_2=c_2(TM)$. They are found
using monad construction. } \label{QuinticBundle}
\end{table}

The tangent bundle is the entry
with rank 3 and $N_{gen}=100$.  In all,
there are five rank-3 and four rank-4 holomorphic vector
bundles that satisfy all the constraints yet yield different
$c_3$. Notice that there are several vector bundles of different
ranks yet with the same Chern classes. It is possible that taking the
direct sum of trivial line bundles with the lower rank vector bundle
and defroming it results in the high rank one.

In general, the monad construction leads to coherent sheaves, not
vector bundles: they have singularities
where the rank of the bundle at that point is smaller than the
generic rank, which happens when the
dimension of the kernel of the map $\phi$ in (\ref{eq:MONAD})
changes.  It is not in general known which singularities are allowed
in the perturbative heterotic string, so to be sure of having examples,
we should check that some of these can be bundles.

Further, we need supersymmetric vacua. This is equivalent to require
the holomorphic vector bundle to be (semi-)stable. In general, it is
not easy
to verify the (semi-)stability of a given holomorphic vector bundle
on a compact manifold, especially when the rank is larger that two.
One must find all the holomorphic subsheaves,
and then compare the slopes $c_1(V)/rank(V)$.

In appendix A, we do both checks for the
holomorphic vector bundles
$V_{(1,1,1,1,2,2;4,4)}$ on the quintic.
This case is interesting bcause it has the same rank as
the direct sum of holomorphic tangent bundle with the trivial line
bundle on the quintic,
yet yields a different number of chiral fermions.
We indeed prove these include vector bundles, but check
stability only against
subsheaves that have a similar monad description, i.e. as a kernel
of a map between direct sum of line bundles on $P^4$. It would
be nice to complete this proof of stability.

Thus there are different embeddings of the $SO(10)$ gauge group in
$E_8$ with different number of generations.  By our general arguments,
these two theories are connected by passing through nonsupersymmetric
vacua.  The question of whether they can be connected through
supersymmetric vacua appears to depend on the precise type of
singularity we allow in the bundle; we leave this for future work.

\section{Smooth chirality change in type IIA orientifold}

In this section, we will give an explicit example in the
context of type IIA orientifold compactification with D6 branes
intersecting at angles. These also give string vacua with $D=4
 \mathcal{N}=1$ together with various gauge groups and chiral fermions. We
construct examples of chirality change through supersymmetric
deformation of brane configurations. The geometric picture makes
the process more explicit. The formalism goes as follows. The
topological constraint satisfied by the homology classes can be
realized by different set of branes wrapping Lagrangian 3-cycles.
Isomorphic effective theory gauge groups can appear with different
content of chiral fermions, naturally arises as different
microscopic brane wrapping in the internal space.

We first briefly review the general
setup of type IIA orientifold compactification, in particular pay
attention to the supersymmetric condition of brane configuration.
Then we construct an explicit example, in the case of
$T^6/Z_2\times Z_2$, of string theory vacua with isomorphic gauge
groups yet different number of chiral multiplets. Further, by a
chain of supersymmetric deformations, we are able to prove that
they are smoothly connected. Finally, we compare with the
heterotic string compactification.

\subsection{Type IIA orientifold compactification on $T^6/\BZ_2\times\BZ_2$}

Type IIA compactification on orientifolds has attracted a lot of
attention recently. The anomaly cancellation requires the
existence of background D-branes to cancel the negative tension
and the R-R charge carried by the orientifold planes. Early work
in $D=6$ compactification see ~\cite{Berkoozetal, BGK}. This
results in the so-called brane-world theory, where both gauge
fields and charged chiral multiplets are present, and, other than
the traditional heterotic/type-I string compactification on
Calabi-Yau, provides another way to realize string vacuum with
phenomenology interests.

In this setup, choose the orientifold $Z_2$ action to be complex
conjugation on the internal Calabi-Yau 3-fold, the orientifold
plane is a real dimension three special Lagrangian submanifold. D6
branes wrapping special Lagrangian submanifold of the internal
Calabi-Yau while extending in all four noncompact dimensions,
results in string theory vacua with d=4 $\mathcal{N}=1$
supersymmetry. If all the D-branes lie within the orientifold
planes, then only symplectic groups appear. However, supersymmetry
allows D-branes to intersect with angles, then unitary groups can
be obtained and chiral fermions \cite{BGKa, FHS} are supported at
the intersection points, finally standard model like spectrum can
be obtained. The simplest geometry allowing chiral theory is
$T^6/\BZ_2\times \BZ_2$. Some of the recent discussions on general
orientifold of the Calabi-Yau manifold can be found in
\cite{ChiralOrient, SMQuintic, OrientMirror, GM1, GM2, B,
  OrientGepner, BW}. Actually the anomaly cancellations conditions allow many
solutions, with various gauge groups and charged matters. This
class of models has been exploited extensively in an effort to
find the standard-model-like spectrum for phenomenological model
building (see \cite{CLSYukawa,CPS}).

In the following, we will concentrate on the orientifold
$T^6/\BZ_2\times \BZ_2$. The orientifold group is generated by
three mutually commutative order 2 group elements, $\{\Omega R,
\theta, \omega \}$. $\theta$ and $\omega$ generate $\BZ_2\times
\BZ_2$ orbifold group, $\Omega$ is a world-sheet parity
transformation while R is a spacetime symmetry acting as complex
conjugation on all three complex coordinates and reversing the
orientation. Use complex variables $(z_1, z_2, z_3)$ to
parameterize the tori, these generators act as follows
\begin{eqnarray}
R&:& (z_1, z_2, z_3) \rightarrow (\bar{z}_1, \bar{z}_2,
\bar{z}_3), \nonumber \\
\theta&:& (z_1, z_2, z_3) \rightarrow (-z_1, -z_2, z_3), \\
\omega&:& (z_1, z_2, z_3) \rightarrow (z_1, -z_2, -z_3). \nonumber
\end{eqnarray}
Quite differently from the two generators of the orbifold group
which preserve the complex structure of the tori, $\Omega R$ is
non-holomorphic and, together with the orientation reversion on
the string worldsheet, it reduces the amount of supersymmetry into
half. When lift to the action on the Chan-Paton factors from the
intersecting branes($\Omega$), it requires a projection on the
gauge bundle that amount to a real (pseudoreal) structure and
results in SO and SP gauge groups when the branes coincide with
the orientifold planes. Notice that each element of the form
$\Omega Rg$ for $g\in\BZ_2\times \BZ_2$ is of order 2, so the
fixed hyperplane under these elements are all special Lagrangian
submanifolds.

The orientifold string theory background has exact conformal field
theory description, despite the presence of the singularities of
the fixed tori and 64 fixed points. It is a particular point of
the moduli space of Calabi-Yau space, viewed as a fibration of K3
space over $P^1$. As shown in \cite{VW}, without discrete torsion,
it belongs to a family of Calabi-Yau manifolds with $h^{1,1}=51$
and $h^{2,1}=3$. Turning on discrete torsion, it can be deformed
into smooth Calabi-Yau manifold with $h^{1,1}=3$ and $h^{2,1}=51$
which is actually the mirror of the Calabi-Yau without the
discrete torsion. For simplicity, we will fix the complex moduli
space of the Calabi-Yau moduli space at point where $T^6$ is a
direct product of complex 2-tori, each fixed at modulus $\tau=i$
(so the lattice is rectangular). The discussion will be similar
for other complex moduli. Throughout this section, we will be
interested only in the open string moduli from the intersecting D6
branes.

The consistent condition for string vacua on the open string
sector is purely topological, just as the constraint on the second
Chern class in the heterotic/type-I case. As usual, this can be
understood from both worldsheet and the spacetime point of view.
From string worldsheet, there is untwisted RR tadpole. Its
cancellation requires the charges of D6-branes and O6-branes to
cancel \cite{BGK,BGKa,FHS}. From spacetime point of view,
orientifold fixed planes carries negative tension and RR charges,
the latter being the source of the RR 7-form. In a compact space,
the total charge have to be zero, which requires to add D6 branes
to cancel. The twisted tadpole can be cancelled by choosing
appropriate projective representations of orientifold group action
on the gauge bundle formed by the overlapping D-branes, as in
\cite{CSU}.

All D6 and O6 branes wrap 3-cycles of $T^6/(Z_2\times Z_2)$. By
supersymmetry, these are calibrated submanifold with respect to
the holomorphic 3-forms of the Calabi-Yau space: $\omega|_C=0,
Re\Omega|_C=V|_C$. Here $\omega$ is the Kahler form, and $\Omega$
is the holomorphic 3-form, not to be confused with the orientifold
action. By definition, these are special Lagrangian submanifold.
And all span the noncompact four dimensional spacetime. In the
following, we will discuss the brane configuration in the covering
space $T^6$. Our example in the following sections will explicitly
contain branes in groups of 4, so one does not need to worry about
whether the branes involved are fractional and have to be stuck to
the fixed point/tori of the orbifold group. On the other hand, the
complex conjugation $R$ is different from element of $\BZ_2\times
\BZ_2$, in that the Calabi-Yau space is not quotient by $R$, but
merely the $\Omega R$ image of each brane is required to be
explicitly included in the background, for example, in the
counting of the homology classes of the D-branes. But in any case,
our example has all the brane images under the whole orientifold
group action.

One of the easiest brane configuration is that they are in
factorized form: each brane is direct product of three circles,
each is a one cycle on the three tori. However, it is important to
realize that these are not the only supersymmetry configuration,
which will be the basis for our discussion of the brane
deformation in the section below. We will fix some notation here.
Let the two fundamental one-cycles of the i-th tori be $[a_i]$ and
$[b_i]$ respectively. Under this basis, any one-cycle on the i-th
torus is $n^i[a_i]+m^i[b_i]$, specified by the winding number
$(n^i, m^i)$. Thus a SLAG with the factorized form has
corresponding homology class
\begin{equation}
[\Pi] = \Pi_{i=1}^3 (n^i[a_i]+m^i[b_i]).
\end{equation}
Its $\Omega R$ image has homology class represented by
\begin{equation}
[\Pi'] = \Pi_{i=1}^3 (n^i[a_i]-m^i[b_i]),
\end{equation}
since $R$ reverses the direction of the b-cycles, while a-cycles
remain invariant. Now one can easily specify the RR tadpole
cancellation condition. For k stacks of D6 branes labelled by
$a=1,...,k$, each stack having $N_a$ branes wrapping a factorized
special Lagrangian submanifold with winding number $(n_a^i,
m_a^i)$, one has
\begin{equation}\label{eq:TADPOLE}
\sum_a N_a[\Pi_a] + \sum_a N_a[\Pi_a'] +(-4)\times 8[\Pi_{O6}] =0.
\end{equation}
Notice that the orientifold branes carry -4 charge each, and
$[\Pi_{O6}]=[\Pi_{\Omega R}]+ [\Pi_{\Omega R\theta}]+[\Pi_{\Omega
R\omega}]+[\Pi_{\Omega R\theta\omega}$ is the total contribution
of all the 3-cycles fixed under group element involving
orientifold action. Similar to the case of heterotic string
compactification, this condition is merely a constraint on the one
particular homology class, and not as constrained as one might
thought. There are many solutions of SLAG's, each realizes a
particular "splitting" of this homology class. We will comment on
this point of view later.

In the factorized form, the supersymmetry condition is easy to
give. Type IIA string theory on Calabi-Yau three-fold has $D=1$
$\mathcal{N}=2$ supersymmetry. The orientifold fixed planes
preserve one particular half of the bulk supersymmetry, determined
by the holomorphic three form. The only way to have
$\mathcal{N}=1$ supersymmetry in $D=4$, is that all the D6-branes
should preserve the same part of the bulk supersymmetry. As
discussed in \cite{BraneSusy}, D6-branes should be related to
O6-branes by a $SU(3)$ rotation. In particular, it allows D6
branes to intersect O6 branes at angles. One of the O6-brane wraps
all the $a$-cycle. Then in the factorized form, assume the one
cycle in the i-th tori wrapped by the brane intersects the real
axis with angle $\theta_i$. Then one has the following
supersymmetric condition
\begin{equation}\label{eq:SusyAngle}
\theta_1+\theta_2+\theta_3=0.
\end{equation}
Branes at generic positions provides unitary gauge group and
possible chiral fields.

We will be concerned with the deformation of branes and the
corresponding change of chiral spectrum. Since we will not
consider the fractional branes, the image branes for a brane at
generic position under the full orientifold group action will be
included in the covering space $T^6$, in particular the $\Omega R$
image. The gauge groups arise as follows. $N$ D6-branes parallel
to orientifold fixed planes produce $USp(N)$ group, while in
generic position with respect to all the orientifold fixed planes,
$U(N)$ group is produced. The orbifold group action reduces it to
$U(N/2)$.

The charged chiral multiplets arise from the intersection of the
various D6-branes, and carry bifundamental representations of the
gauge groups from the intersecting branes. A necessary condition
for chiral fermions in d=4 is that they carry complex
representation of the gauge groups. Therefore it has to involve at
least one $U(N/2)$ group (there is no possibility of spinor
representation of the $SO(N)$ in this scenario). For $2N_a$
D6-branes intersecting $2N_b$ branes, there are $I_{ab}$ chiral
fermions in $(N_a, \bar{N}_b)$ from $(ab+ba)$ sector, and
$I_{ab'}$ chiral fermions from $(ab'+b'a)$ sector, where b' is the
$\Omega R$ image of b. The intersection number $I_{ab}$ is defined
as
\begin{equation}\label{eq:InterNumb}
I_{ab}=[\Pi_a]\cdot [\Pi_b] = \Pi_{i=3}^3 (n_a^im_b^i-m_a^in_b^i).
\end{equation}
Apart from above fields, there are three $\mathcal{N}=1$ chiral
multiplets in the adjoint representation associate with each stack
of D6-branes, result from spontaneous breaking of $\mathcal{N}=2$
supersymmetry. Chiral multiplet in the antisymmetric
representation of $U(N)$ also arises from $2N$ D6 branes and its
$\Omega R$ image brane. These will parameterize the brane
recombination process, which is important for chirality change
process we will propose in the following section.

\subsection{Smooth brane deformation and supersymmetric chirality change}

Let us look more carefully at the RR tadpole cancellation
constraint eqn.(\ref{eq:TADPOLE}). Expanded out explicitly in the
homology basis, it takes the following form
\begin{equation}
[a_1] \times [a_2] \times [a_3]: \qquad \sum_a N_a n_a^1 n_a^2 n_a^3= 16, \\
\end{equation}
\begin{equation}
[a_1] \times [b_2] \times [b_3]: \qquad \sum_a N_a m_a^1 m_a^2 n_a^3= -16, \\
\end{equation}
\begin{equation}
[b_1] \times [a_2] \times [b_3]: \qquad \sum_a N_am_a^1n_a^2m_a^3=-16, \\
\end{equation}
\begin{equation}
[b_1] \times [b_2] \times [a_3]: \qquad \sum_a
N_am_a^1m_a^2n_a^3=-16.
\end{equation}
It is easy to understand why there are four independent equations
here. Recall that the orbifold space has $h^{2,1}=3$, add
$h^{3,0}=1$, so the homology 3-classes is of complex dimension
four. Orientifold action acts as complex conjugation, so its
invariant set within the homology classes is of real four
dimensional. However, the D6 branes can wrap homology three cycles
containing homology classes in the other four dimensional space.
Since they are odd under complex conjugation, when the brane and
its $\Omega R$ image add together, the components along these
direction simply cancel out. This is one hint why the constraint
(\ref{eq:TADPOLE}) is not very restrictive: it is constraint on
homology classes, not on actual 3-cycles.

Now it is a simple matter to find different solutions to this
constraint equation. Many can be found, as in \cite{CSU}. Each has
different gauge groups and matter content. Our theme in this paper
is to find classical process of chirality change, so we would like to
know first if there are solutions with isomorphic low energy gauge
groups, but different chiral fermions. Indeed, there are examples
within this setup, which is listed in Table~\ref{BraneConfig}.

\begin{table} \footnotesize
\begin{center}
\begin{tabular}{||c|c||c|c||c|c||c|c||}
  \hline
  \hline
  \multicolumn{2}{||c||}{Brane type} & \multicolumn{2}{|c||}{Phase I}
  & \multicolumn{2}{|c||}{Phase II} & \multicolumn{2}{|c||}{Phase III}\\
  \hline
   a & $\Pi_i(n_a^i, m_a^i)$ & $N_a$ & gauge group & $N_a$ & gauge
  group & $N_a$ & gauge group \\
    \hline \hline
  A & $(1,0)\times(1,1)\times(1,-1)$ & 4 & $U(2)$ & 4 & $U(2)$ & 4 & $U(2)$ \\
    \hline
  B & $(1,1)\times(1,-1)\times(1,0)$ & 12 & $U(6)$ & 0 & & 0 & \\
  \hline
  C & $ (1,1)\times (1,0)\times(1,-1)$ & 0 & & 0 & & 12 & $U(6)$ \\
  \hline
  D  & $(1,0)\times(1,0)\times(1,0)$ & 0  &  & 12 & $USp(12)$ & 0 &  \\
\hline
  E  & $(1,0)\times(0,1)\times(0,-1)$ & 12 & $USp(12)$ & 12 &
  $USp(12)$ & 12 & $USp(12)$ \\
\hline
  F  & $(0,1)\times(1,0)\times(0,-1)$ & 16 & $USp(16)$ & 16 &
  $USp(16)$ & 4 & $USp(4)$ \\
\hline
  G  & $(0,1)\times(0,-1)\times(1,0)$ & 4 & $USp(4)$ &  $12+4$ &
  $USp(16)$ & 16 & $USp(16)$ \\
\hline\hline
\end{tabular}
\end{center}
\caption{\small D6-brane configurations on covering space $T^6$
and corresponding gauge groups. I and III have isomorphic gauge
groups but different chiral multiplets. They are proposed to
connect in a supersymmetric way, via configuration II. }
\label{BraneConfig}
\end{table}

Three brane configurations are given in the table, which are labelled as I, II
and III, together with the gauge groups. Altogether there are
seven types of branes (or rather 3-cycles) involved, listed as
type A,..., G. The first three types of 3-cycles do not coincide
with any orientifold planes and so produce unitary groups, while the
last four types coincide with the orientifold plane and produces
symplectic gauge group.

Notice I and III have isomorphic low energy gauge groups, which is
$U(2)\times U(6) \times USp(12) \times USp(16) \times USp(4)$. Now
it is clear, that these D6-branes wrap different 3-cycles in
the internal Calabi-Yau space, so indeed the gauge groups are
different but merely isomorphic. In particular, the chiral
spectrum are different. The intersection number of the various
branes can be easily worked out from (\ref{eq:InterNumb}),
\begin{eqnarray*}
   I_{AB} = -2, I_{AD} &=& 0, I_{AE}=0, I_{AF}=1, I_{AG}=-1;\\
   I_{BD} = 0,  I_{BE} &=& 1, I_{BF}=-1, I_{BG}=0; \\
   I_{AC} = 0, I_{CD} &=& 0, I_{CE}=1, I_{CF}=0, I_{CG}=-1.
\end{eqnarray*}
Then the the bifundamental chiral multiplets from the D6-branes
intersecting points in phase I and III are 
\begin{eqnarray*}
U(2) &\times & U(6) \times USp(12) \times USp(16) \times USp(4) \\
I:-2(2,\bar{6}, 1,1,1) &+& (2,1,1,16,1)
-(2,1,1,1,4)+(1,6,12,1,1)-(1,6,1,16,1).  \\
III: &-&(2,1,1,16,1)+(2,1,1,1,4) + (1,6,12,1,1) - (1,6, 1,16,1).
\end{eqnarray*}
The minus sign denotes the open string with opposite direction,
therefore the corresponding chiral multiplets carry complex
conjugate representations with respect to the one with positive
sign. For example, $-2(2,\bar{6}, 1,1,1)$ is equivalent to 2
chiral multiplets carrying bifundamental representation
$(\bar{2},6)$ under $U(2)\times U(6)$, and being singlet under all the
other gauge groups. Again, we have omitted the
non-chiral spectrum arising from open string between brane and its
$\Omega R$ image, and between a brane with itself.

Evidently, there is a nontrivial change of chirality between I and
III. The two chiral multiplets carrying bifundamental
representation of $U(2)\times U(6)$ in I disappear from III, while
there is more subtle change of representation from $(2,16)$ under
$U(2)\times USp(16)$ in I to $(\bar{2},16)$ in III, and from
$(2,4)$ under $U(2)\times USp(16)$ in I to $(\bar{2},4)$ in III.
If we ignore the spectator branes in these two configurations, the
change can be described in terms of brane types as
\begin{equation}\label{eq:OneToThree}
 12B+ 12B' + 24F \rightarrow 12C+12C' +24 G.
\end{equation}
The F and G branes have double number than those listed in the
table, because their $\Omega R$ image branes are themselves, which
are also included in the equation.

One would like to know how configurations I and III are connected.
We propose that they can be connected through configuration II.
The gauge groups and chiral spectrum of II are as follows,
\begin{eqnarray*}
II: U(2)\times Usp(12) &\times& USp(12) \times USp(16) \times USp(16) \\
(2,1,1,16,1) &-& (2,1,1,1,16).
\end{eqnarray*}

One realization of the chirality change
(\ref{eq:OneToThree}) proceeds in two steps as
follows. First, from I to II
\begin{equation}\label{eq:OneToTwo}
    12B+ 12 B' \rightarrow 24 D+24 G.
\end{equation}
while the second step is from II to III
\begin{equation}\label{eq:TwoToThree}
    24 D+ 24 F \rightarrow 12C+ 12 C'.
\end{equation}
Each only involves the brane configuration in a $T^4$, and all the
branes involved in one process wrap the common one-cycle in the
other torus. Via the above brane ``chemical reaction,"
the chiral spectrum is changed.

\subsubsection*{Brane supersymmetric deformation}

Having narrowed down the deformation to a more specific process,
one wonders whether this can be realized in a supersymmetric way.
At first, the supersymmetric condition (\ref{eq:SusyAngle}) on
the angles of the brane seems very constrained. For example, in
(\ref{eq:TwoToThree}), $\theta_3=0$ is fixed, which leaves only one
deformation parameter, say $\theta_1$, from solving the
constraint. On the first $T^2$, because of the periodicity of the
torus, $\theta_1$ can only take discrete values, $\theta
=\tan^{-1} (m/n)$. This might suggest that the
supersymmetric solutions are isolated ``islands" in the configuration
space, and no supersymmetric deformation can connect them.

However it is simple to see that at least some supersymmetric
deformations are possible.  The McLean theorem\cite{McL} states that
for a compact special Lagrangian submanifold $N$ in a Calabi-Yau
manifold, the moduli space of special Lagrangian deformation of $N$ is
a smooth manifold of dimension $b^1(N)$, where $b^1(N)$ is the first
Betti number of $N$. A heuristic argument for this uses supersymmetry
and world-volume gauge theory. Assume a D-brane wraps $N$, which has a
$U(1)$ bundle on it. The flat Wilson lines then parameterize the
moduli space of dimension $b^1(N)$. Paired with another moduli from
minimal area of the string worldsheet in the Calabi-Yau with boundary
on the SLAG wrapping the 1-cycle, they form a
complex parameter.

Now the orbifold $T^6/\BZ_2\times \BZ_2$ is simply connected. This is
clearly the case as all the one-cycles in the original $T^6$ are
odd under a certain orbifold group element, and so all disappear
after the orbifold projection. However all the special Lagrangian
3-cycle $C$ discussed so far has the topology of the tori of real
dimension three, and invariant under the whole orbifold group.
This means that in the orbifold such a SLAG has moduli space of
complex dimension three. A single wrapped D-brane must also have moduli
space of complex dimension three. If multiple branes wrap the same
SLAG submanifold, the moduli space will increase correspondingly,
as the Wilson lines can take value in $U(n)$ gauge group now.

The consideration of multiple wrapped branes brings us to another
important point. In the case at hand, one encounters not merely a
single brane wrapping a SLAG, as one notices that there are large
number of branes wrapping the same SLAG cycle in the solutions to
the constraint equation given above. In addition, a brane's
$\Omega R$ image is always present because of the orientifold
action. Thus the object to deform, is not certain singly wrapped
SLAG submanifold, but the combined 3-cycle with different
components wrapped by multiple branes. In particular, splitting
and recombining are possible in this setup.

Let us now describe the supersymmetric deformation from I to II
(\ref{eq:OneToTwo}). First observe that the branes
involved (B, D and G in our notation) all wrap the same $a$-cycle
in the third torus, i.e. $\theta_3=0$, and thus remains the same
though the deformation. So the deformation actually happens in
four-dimensional complex space $\cal{T}\times \cal{T}$, and
original special Lagrangian 3-cycles reduce to SLAG 2-cycles. The
advantage is that algebraic geometry technique can be applied, since SLAG in
4d space is just complex curve with respect to a different complex
structure, as shown in \cite{BraneSusy, SLAG}, where the
coordinate system is 
\begin{equation}
w_1=x_1+ix_2,\qquad w_2=y_1-iy_2.
\end{equation}

Now it is easy to identify B, B', D and G type branes with complex
curves in the new coordinate system, as
\begin{equation}
w_1=\pm w_2 \qquad (B, B'); \qquad w_2=0 \qquad(D); \qquad w_1=0
\qquad (G).
\end{equation}
The orientifold action $\Omega R$ acts as $(w_1, w_2)\rightarrow
(w_1, -w_2)$ in the new coordinate system.

\begin{figure}
\begin{center}
\centering \epsfysize=3.5cm \leavevmode
\epsfbox{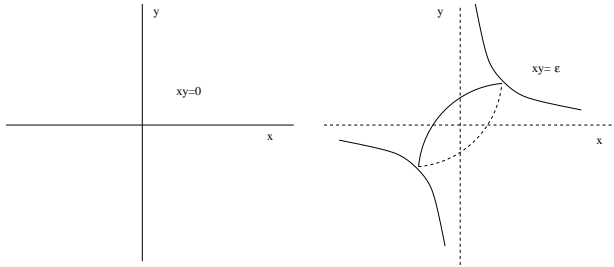}
\end{center}
\caption[]{\small Two intersecting complex curves described by the
  equation $xy=0$ deform to a a configuration described by the equation
$xy=\epsilon$. x and y are complex coordinates, so these are Riemann
  surfaces, and the one in the second picture is
  connected. It would be easier to visulize if one regards this as
  degree-2 curve defined in a coordinate patch specified by $z\neq 0$ in
  $P^2$ with homogeneous coordinates (x,y,z). }
\label{braneB}
\end{figure}

Having realized the non-trivial part of the brane configuration as
complex curves, the deformation is straightforward.  The local picture
is illustrated by figure \ref{braneB}\cite{CallanM}. This picture of
brane recombination, while preserving supersymmetry at
every stage, is analyzed in detail in \cite{EGHK} for the noncompact
target space, which can be viewed as the local model for the process
we describe here.  

To get a global description, we realize the complex curve as complex
intersection of the 4-tori $T^4$ with a hypersurface in the
product of the projective space $P^2\times P^2$, deforming the
hypersurface will results in a family of holomorphic curves that
connects the two configurations. Each $T^2$ is nonsingular, can be
embedded as cubic curve in $P^2$, which in terms of the
homogeneous coordinates $(x_i,y_i,z_i)$ on $P^2$ is described by
\begin{equation}\label{eq:ToriEmbed}
y_i^2z_i=4x_i^3-g_2x_iz_i^2-g_3z_i^3, \qquad i=1,2.
\end{equation}
These equations define separately a bidegree $(3,0)$ and a $(0,3)$
divisor in $P^2\times P^2$, which are topologically $T^2\times
P^2$ and $P^2\times T^2$. Their intersection will produce
$T^2\times T^2$. Notice that the two tori has the same complex
structure, so $g_2$ and $g_3$ are the same for the two factors.

The supersymmetric $B, B', D, G$ branes are complex curves in
$T^4$. We now find appropriate divisors in $P^2\times P^2$, so
that their intersection with the two divisors \ref{eq:ToriEmbed}
defining the two embedded tori will give these curves. For this
purpose, we need the map from the tori to the homogeneous
coordinates, which is through Weierstrass $\wp$-function,
\begin{eqnarray}
x_i &=& \wp(w_i) = \frac{1}{w_i^2} +\sum_{0\neq\lambda\in\Lambda}
[\frac{1}{(w_i-\lambda)^2}-\frac{1}{\lambda^2}], \\
y_i &=& \wp'(w_i) = \sum_{\lambda\in\Lambda}
\frac{-2}{(w_i-\lambda)^3}, \qquad i=1,2,
\end{eqnarray}
where the lattice $\Lambda=\{m+n\tau|m,n\in \BZ\}$ has period
determined by the complex structure $\tau$ of the tori, which is
$\tau=i$ for the case at hand.

For the D and G branes, $w_2=0$ and $w_1=0$ are mapped to
$x_1=y_1=\infty$ and $x_2=y_2=\infty$ respectively from
Weierstrass function mapping in the coordinate patch $z=1$. Thus
they can be identified with equation $z_2=0$ and $z_1=0$
respectively. There are complications from the degree counting in
determining the actual number of wrapped D-branes. An elliptic
curve has degree-3 in $P^2$, which intersects degree-1 divisor
$z_1=0$ at three points. Further intersecting with the torus in
the second $P^2$, it produces 3 copies of the second torus, which
is equivalent to three copies of G-type branes. Similarly, $z_2=0$
corresponds to three copies of D-type branes. This is the common
feature of embedding tori by Weierstrass form: its intersection
with any divisor will be multiple of three. We have arranged our
brane solution meet this requirement. Excluding the common factor
of 4 for the possible orbifold images, the brane combination
$24D+24G$ is described by intersection of the tori with the
divisor
\begin{equation}\label{eq:DivOne}
z_1^2z_2^2=0.
\end{equation}

The $B, B'$ branes equations are $w_1\pm w_2=0$. Now under mapping
by Weierstrass function, it leads to $x_1=x_2$ and $y_1\pm y_2=0$
in $P^2\times P^2$. Account for the degree, $3B+3B'$ in the
coordinate patch $z_1=z_2=1$ is  $y_1^2-y_2^2=0$. In homogeneous
coordinates, this leads to equation
\begin{equation}\label{eq:DivTwo}
y_1^2z_2^2-y_2^2z_1^2=0.
\end{equation}
Because $\Omega R$ acts as $y_2\rightarrow -y_2$, so above
equation is invariant under $\Omega R$, as it should.

Both (\ref{eq:DivOne}) and (\ref{eq:DivTwo}) are degree $(2,2)$
divisors on $P^2\times P^2$. Orientifold action $\Omega R$ acts on
the second torus only, $\Omega R: w_2\rightarrow -w_2$, which can
be lifted to $P^2$ as $\Omega R: y_2\rightarrow -y_2$. So if the
defining polynomial is even function of $y_2$, it is automatically
$\Omega R$ invariant. Thus the family of hypersurface which both
(\ref{eq:DivOne}) and (\ref{eq:DivTwo}) belong to can be written
as
\begin{equation}
F_{(2,2)}(x_1,y_1,z_1,x_2,y_2^2, z_2)=0.
\end{equation}
Deformation within this family can make a transition from
configuration I to II.

The intersection of (\ref{eq:DivOne}) and (\ref{eq:DivTwo}) with
$T^2\times T^2$ each has the topology of two torus intersecting at
one point However the generic member of the family of the $(2,2)$
divisor intersecting with $T^2\times T^2$ has quite different
topology. In the notation of \cite{H}, this is specified by a
configuration matrix 
\begin{equation}
\left(\begin{matrix} 2 \\ 2 \end{matrix} \right|\left|
\begin{matrix} 3 & 0 & 2 \\ 0 & 3 & 2 \end{matrix} \right).
\end{equation}
where the first column specifies the dimensions of the product
projective spaces, and the other columns specify the degrees of
the divisors corresponding to each projective space. The configuration
represents a family of
complete intersections (can be smooth or singular) defined by
three polynomial constraints specified by the configuration
matrix. It is parameterized by the space of coefficients of the
polynomials. A generic member of this family is a smooth complex
curve. Let us calculate the genus of this curve $C$.

Let the K\"{a}hler forms of the two $P^2$ be $J_1$ and $J_2$. The
tangent bundle of $C$ is a sub-bundle of $TX$, $X\equiv P^2\times
P^2$, whose quotient is the vector bundle $E$ determined by the
polynomials. In other words, there is an exact sequence 
\begin{equation}
0\rightarrow TC \rightarrow TX|C \rightarrow E|_C \rightarrow 0.
\end{equation}
This implies the relation of the total Chern class
\begin{equation}
c(C) = \frac{c(X)}{c(E)}
=\frac{(1+J_1)^3(1+J_2)^3}{(1+3J_1)(1+3J_2)(1+2J_1+2J_2)} =
-2(J_1+J_2).
\end{equation}
The Euler character is obtained from the coefficient of the volume
form $J_1^2J_2^2$ of $P^2\times P^2$, in the product
$$
c_1(C)\wedge ((1+3J_1)(1+3J_2)(1+2J_1+2J_2)),
$$
which gives $\chi(C)=-72$. By the familiar Riemann-Hurwitz formula
for curves $\chi(C)=2-2g$, this leads to genus of $C$ being 37.

Now it is clear that the two configurations that we tried to
connect holomorphically are really two degenerate limits of the
same family of curves with genus $g=37$. These degenerate
configurations develop
intersecting points which are singular. But these singular limits have a
perfect classical interpretation in the gauge theory, that is the
brane splits into
multiple branes wrapping different supersymmetric cycles, together
with intersections where chiral multiplets live.

As a last step of the proof of the supersymmetric transition
realizing chirality change, one notice that the transition from II to III
(\ref{eq:TwoToThree}) can be similarly proved. Switching the role of
the second torus with the third torus, it becomes the same process
as the transition from I to II.

Thus, we have an explicit example, in which by recombination and
splitting, chirality change with isomorphic gauge groups takes place
entirely in the moduli space of supersymmetric string vacua.

\section{Conclusions}

We showed that string theory compactifications leading to low energy
effective field theories with different chiral matter content are
connected even at the classical level, in a way which can be
understood from several viewpoints: effective field theory, higher
dimensional gauge theory and brane constructions.

The loophole in the naive argument that this is not possible
is that the true effective field theory description of the
situation is in terms of an infinite gauge group, which in the
gauge theory examples is the group of all maps from the internal
space to the (finite dimensional) gauge group, which contains many
copies of any given low energy gauge group.  Chirality change
takes place by breaking one copy of the low energy gauge group,
and restoring a different copy with different chiral matter.

We even found an example in which the transition takes place
entirely on the supersymmetric moduli space, in \IIa\ theory
with D6-branes.  Again, there is no particular paradox in this.
In a partially understood sense, this is mirror to similar transitions
between type \I\ vacua (via T-duality/mirror symmetry) and the
heterotic string (via S-duality).  On the other hand, detailed
understanding of the mirror map usually involves $\alpha'$ corrections,
leaving open the possibility that these corrections are required to
get transitions purely through supersymmetric classical vacua.
It would be nice to find an example of a transition
purely in ten dimensional Yang-Mills theory (we suspect it is possible).

The larger question which this investigation set out to address is
to what extent we can think of ``all'' or large
sets of string theory compactifications as described by a single
effective field theory, the ``effective field theory of
everything,'' with all quantum effects already summarized in the
effective Lagrangian. Such an effective field theory would have to
contain an infinite number of fields, almost all of which are
massive in any small region of configuration space, and would
probably not be the best way to study any one compactification.
Nevertheless it might be very useful as a unifying and organizing
concept behind the general discussion.

One can try to propose specific formal realizations of this idea
such as the explicit Kaluza-Klein expansion of higher dimensional
gauge theory keeping all modes which can become massless anywhere
in the region of configuration space of interest, a similar
discussion based on world sheet considerations along the lines
proposed for toroidal compactification of the heterotic string by
Giveon, Porrati and Rabinovici \cite{GPR}, a similar proposal based
on string field theory, or otherwise.

However, the main question which will determine whether there is
an effective field theory of everything or not, is not a formal
question. Rather, it is the physical question of what type of
phase transitions connect the different vacua of string/M theory.
Certain phase transitions can be easily described in effective
field theory language, such as the Coulomb-Higgs transition.
Others can be so described by taking advantage of duality, such as
the Coulomb-confinement transition.  While going beyond the
traditional idea of effective field theory, a collection of
effective field theories related by explicit dualities, with
different theories appropriate in different regions of
configuration space, would be a pretty concrete description of the
situation.

While such a picture is often cited as the basic picture behind
string/M theory duality, it is not yet clear whether one can use
it in this literal form, as there is no guarantee that all of the
phase transitions one needs to connect the different vacua are of
this form. For example, there is no clear argument in supergravity
that first order transitions are impossible (indeed one cannot
take for granted standard considerations of minimizing the
energy).

Another example is a phase transition associated with a
non-trivial fixed point theory. While one should certainly not try
to describe the physics of the phase transition itself by
effective field theory, one might hope that there exists an
effective field theory which describes all massive deformations of
the fixed point theory and the configuration space in its
vicinity.
This is problematic in the case of a non-trivial fixed point theory which
arises at the meeting point of several branches of moduli space, which
individually are described by very different sets of degrees of
freedom.  The simplest effective field theory description would have
to contain all of these degrees of freedom, and on each branch give
mass to all fields associated with other branches. This is impossible
if some of these theories are chiral.  One might gain more
possibilities by postulating dualities between some of these fields
and others, but in any case there are many examples for which such a
description is not known at present, and might be impossible.

Our present discussion suggests that, even in cases where this is
impossible, it might still be possible to describe all of the branches
within a single effective field theory, which connects them up along
a different line of deformations.  It would not be manifest that
they are also connected through non-trivial phase transitions, but
this information might be taken into account separately.

One approach to show that this is impossible, would be to find an
order parameter which could not change in effective field theory.
In the case at hand, we showed that the number of generations is not
the relevant order parameter.  On the other hand, there is a
candidate order parameter in the heterotic string construction. It is
$c_2(V)$ or equivalently the number of heterotic fivebranes.  Thus, to
judge the connectedness of this moduli space in effective field
theory, we must ask, is $c_2(V)$ preserved under all processes which
do not pass through non-trivial fixed point theories?  While true in
the heterotic string, it is not at all obvious in the full
string/M theory, again because there are dual descriptions, such as
$G_2$ compactification, in which it is not obvious how to define this
number.

As a final comment, all of the connections we have been talking about
between solutions of string theory are visible at low energy.  A
possibly more important question is whether there are simpler ways to connect
solutions in a regime of large quantum and thermal fluctuations, such
as one expects to describe early cosmology.  Though there is
interesting work on the related question of time dependent
interpolating solutions (e.g. describing brane-antibrane
annihilation), it seems to us that at present nothing is known about
this last topic.  Perhaps the concrete pictures we have explored
in this work could inspire some good questions here.

\medskip

We thank B. Acharya, F. Bogomolov, A. Langer, N. Seiberg and R. Thomas
for valuable discussions.

M.R.D. is supported by a Gordon Moore Visiting Professorship at Caltech,
and by DOE grant DE-FG02-96ER40959. C-G.Zhou is supported by DOE grant DE-FG01-00ER45832.

\section*{Appendix A Vector bundles $V_{(1,1,1,1,2,2;4,4)}$ on quintic Calabi-Yau}

In this appendix, we prove the proposition in section 3.2
concerning properties of the bundle $V_{(1,1,1,1,2,2;4,4)}$ on
quintic Calabi-Yau.

First, as demonstrated in section 3.2, this bundle has Chern
classes
\[ r=4, c_1=0, c_2= 50, c_3= 90.\]
Obvious it is not in the complex moduli space of $TM\oplus
\mathcal{O}$. The complex moduli space is parameterized by the map
from $\mathcal{O}(1)^{\oplus 4}\oplus\mathcal{O}(2)^{\oplus 2}$ to
$\mathcal{O}(4)^{\oplus 2}$. Explicitly,
\[ F= \begin{pmatrix} f_1 & f_2 & \cdots & f_6 \\
g_1 & g_2 & \cdots & g_6 \end{pmatrix}, \] where $f_1,\cdots f_4$
and $f_1,\cdots f_4$ are homogeneous degree 3 polynomials in
homogeneous coordinates $X_1, X_2, \cdots, X_5$ of $P^4$, and
$f_5, f_6, g_5, g_6$ are homogeneous degree 2 polynomials.

We first discuss the question whether generically V is a vector
bundle. This is important for interpretation of this vector bundle
as a perturbative heterotic string vacuum. The reason is that
singular fibers corresponds to low dimensional branes as small
instanton limit of the vector bundle, which signals the existence
of the non-perturbative degrees of freedom.

To be rid of singular fibers, the rank of the fibers of V must be
constant over all of the Calabi-Yau space. Because V is defined as
the kernel of the map $F$, this is equivalent to the image of $F$
should have constant rank 2 at each point of Calabi-Yau. Actually
for $V_{(1,1,1,1,2,2;4,4)}$ on quintic, we have a stronger result,
namely that V is generically a vector bundle on the whole $P^4$,
not merely on the quintic hypersurface.

The proof goes as follows. Assume at point $x_0\in P^4$, $dim
(Im(F)) <2$. This means that the six two-dimensional vectors
$\binom{f_i}{g_i}|_{x_0}, i=1,2,\cdots, 6$ are proportional to
each other. This is true if and only if the five functions
$det\begin{pmatrix}f_1 & f_i \\ g_1 & g_i \end{pmatrix},
i=2,\cdots, 6$ are all equal to zero at $x_0$. So the position of
singular fibers are the solutions to the 5 equations
$f_1g_i-g_1f_i=0, i=2, 3, \cdots, 6$. Generically, there is no
solution to these equations as five divisors have no intersection
on $P^4$ generically. Thus we can always find the map $F$ such
that V is a vector bundle.

Next we discuss the stability property of $V_{(1,1,1,1,2,2;4,4)}$.
We would like to know whether there exists a stable vector bundle
in the same complex deformation class of V. This is again by
analysis of F. We can only obtain partial results, namely, we can
prove that if choosing appropriately, V is stable for a class of
subsheaves which also has monad like description. In general the
stability property of a vector bundle is a very hard problem
practically, especially for high rank vector bundles at high
dimensional space. In view of this difficulty, although we only
can prove the stability partially, we feel that it is a strong
evidence that V can be a N=1 perturbative heterotic string vacuum.

A vector bundle is stable if all its subsheaves $V'$ with strictly
smaller rank satisfies the condition
$\frac{c_1(V')}{rank(V')}<\frac{c_1(V)}{rank(V)}$. Here $\mu(V)=
c_1(V)/rank(V)$ is called the slope of the vector bundle, and the
stability criterion we use here is called slope stability
condition. This is the appropriate stability condition at the
large volume limit of the Calabi-Yau space, otherwise we should
use the $\Pi$-stability suitable for stringy Calabi-Yau.

Any vector bundle $V'$ has injective resolution, which is a complex
\[0\rightarrow V'\rightarrow W_1\stackrel{G}{\rightarrow} W_2 \rightarrow \cdots. \]
whose cohomology is trivial except at the first term. The vector
bundles $W_i$ are injective objects in the derived category of
torsionless sheaves. In case of $P^4$, these injective objects are
simply direct sum of line bundles of various degrees,
$W_i=\oplus_k\mathcal{O}(k)$. For $V'$ to be subsheaf of V means
the following diagram commutes,
\begin{equation*}
\begin{array}{ccccccccc}
0 & \rightarrow & V & \rightarrow & \mathcal{O}(1)^{\oplus 4}\oplus
\mathcal{O}(2)^{\oplus 2} & \stackrel{F}{\rightarrow} &
\mathcal{O}(4)^{\oplus 2} & \rightarrow & 0 \\
&  & \Big\uparrow &  &  \Big\uparrow\phi_1 & & \Big\uparrow\phi_2 & & \Big\uparrow \\
0 & \rightarrow & V' & \rightarrow & W_1 & \stackrel{G}{\rightarrow}
& W_2 & \rightarrow & \cdots \\
&& \Big\uparrow &&&&&& \\
&& 0 &&&&&&
\end{array}
\end{equation*}
On $P^4$, a generic sheaf has an injective resolution with five
terms, by analogous construction of Koszul resolution. Because of
difficulty to list all of these sheaves, we will merely check
those subsheaves of V such that they can be similar described as a
kernel of a surjective map between two direct sum of line bundles.
This is the same as saying V' has the same description as V.

The injective resolution of a sheaf is not unique. We assume a
minimal resolution of V'. This means that we choose $W_i$'s to
have minimal rank among all the equivalent injective resolutions
of V'. This condition, combined with the injection of map
$0\rightarrow V' \rightarrow V$, yields the injection $W_1
\rightarrow \mathcal{O}(1)^{\oplus 4}\oplus\mathcal{O}(2)^{\oplus
2}$. But $\phi_2$ may has a nontrivial kernel, and yet the diagram
still commutes. This causes possible complications.

The potentially destabilizing subsheaf $V'$ satisfies the
following constraints:
\begin{enumerate}
\item $C_1(V')=c_1(W_1)-c_1(W_2)>0$; \item $W_1
\stackrel{\phi_1}{\rightarrow} \mathcal{O}(1)^{\oplus
4}\oplus\mathcal{O}(2)^{\oplus 2}$ is injective; \item $G$ and
$\phi_i$ should all have nonnegative degrees.
\end{enumerate}
There are only a finite number of possibilities. Especially
condition (2) restricts the line bundles in $W_1$ to have degree
2, 1, 0, ....

Case I. $\mathcal{O}(2)\subset W_1$, and this line bundle is
mapped to $W_2$ nontrivially. There are three possibilities. (The
factor $\mathcal{O}(2)$ in $W_2$ does not give anything
nontrivial. $\mathcal{O}(1)$ in $W_2$ is not involved.)
\begin{enumerate}
\item $W_2=\mathcal{O}(3)$. $W_1$ must be equal to one of the
$\mathcal{O}(2)$ factors (assume to be the last factor) after a
suitable linear transformation. Thus we have commuting diagram
\begin{equation*}
\begin{array}{ccccc}
\mathcal{O}(2) & \stackrel{F}{\rightarrow} & \mathcal{O}(4)^{\oplus 2} & \rightarrow & 0 \\
\Big\uparrow\phi_1 & & \Big\uparrow\phi_2 & & \\
\mathcal{O}(2) & \stackrel{G}{\rightarrow} & \mathcal{O}(3) & \rightarrow 0 \\
\end{array}
\end{equation*}
The commuting condition, $\phi_3 \cdot G =F$ is
\[  \binom{(\phi_2)_1}{(\phi_2)_2}G= \binom{f_6}{g_6}, \]
where $(\phi_2)i$ and $G$ are homogeneous degree one polynomials,
and $f_6,g_6$ are homogeneous degree two polynomials. This should
be regarded as an algebraic equation, i.e. we should expand out
both side and compare coefficients of monomials. In general, this
results in a set of quadratic equations for the coefficients of
$(\phi_2)i$ and $G$, and the coefficients of $f_6,g_6$ are given.
Counting the number of unknowns ($3\times 5=15$) and the number of
equations (which is the number of monomials in $f_6$ and $g_6$,
i.e. $2\times 15=30$), it is obviously an over-determined set of
equations. So generically, there is no solution for $\phi_2$ and
$G$.

\item $W_2=\mathcal{O}(3)\oplus \mathcal{O}(3)$. Simple
dimensional counting similar to above case does not work. We have
to consider all the conditions. Including the $c_1(V')$ positive
condition , this occurs only if the resolution of $V'$ is
\[ 0\rightarrow \mathcal{O}(1)^{\oplus 2}\oplus \mathcal{O}(2)^2
\stackrel{G}{\rightarrow} \mathcal{O}(3)^{\oplus 2} \rightarrow 0.
\]
The image of $\phi_1$ can be made in a five dimensional subspace
via a trivial linear transformation, and the only nontrivial part
of $\phi_1$ is $\mathcal{O}(1)^3\rightarrow \mathcal{O}{2}^2$.
Spell out the commuting condition, we have
\[ \begin{pmatrix} (\phi_2)_{11} &  (\phi_2)_{12} \\
  (\phi_2)_{21} &  (\phi_2)_{22} \end{pmatrix}
\cdot \begin{pmatrix} G_{11} &  G_{12} & G_{13} & G_{14} &  G_{15}\\
G_{21} &  G_{22} & G_{23} & G_{24} & G_{25} \end{pmatrix}
= \begin{pmatrix} f_2 &  f_3 & f_4 & f_5 &  f_6\\
 g_2 &  g_3 & g_4 & g_5 & g_6 \end{pmatrix}
 \cdot \phi_1. \]
There are 160 unknowns in the coefficients of $\phi_1, \phi_2$ and
$G$, but there are 270 equations. Excluding the linear
transformation between the equivalent line bundles in the same
sire sum, this reduce the equation to 270. Again, this is an
over-determined set of equations, and generically there is no
solutions.

\item $W_2=0$. This is the same as saying V contains
$\mathcal{O}(2)$ as a sub-bundle. Obviously, a generically F will
involve both $\mathcal{O}(2)$ factors, and there should not be
such nontrivial sub-line bundle n V. Direct proof via dimension
counting is very similar to case one, and we will omit the
details.

\end{enumerate}

Since we require that the resolution of $V'$ is minimal, so if
there is a factor $\mathcal{O}(2)$ in $W_1$ was involved
trivially, there must be a resolution with smaller resolution
without this factor. So we can just consider that $W_1$ only
contains $\mathcal{O}(1)$ factors.

Case II. $W_1=\mathcal{O}(1)^{\oplus m}$, m=1,2,3,4. We have the
following possibilities.
\begin{enumerate}

\item
$W_1=\mathcal{O}(1), W_2=0$. This is similar to case I.3.
There are no solutions generically.

\item $W_1=\mathcal{O}(1)^2$. By a trivial linear transformation
of the two line bundles, this decomposes into direct sum of two
series. No solutions generically.

\item $W_1=\mathcal{O}(1)^3$. The only nontrivial case is
$W_2=\mathcal{O}(2)$. Similar to case I.2. There are 135 unknowns,
but there are 270 equations. Generically, there is no solution.

\item

$W_1=\mathcal{O}(1)^4$. The nontrivial cases are
$W_2=\mathcal{O}(2)$ or $\mathcal{O}(3)$.Similar to case I.2.
There are 170 and 110 unknowns respectively,, and 340 equations
for both case. Again there is no solutions generically.

\end{enumerate}

All the other possibilities will involve the cases considered
above. We can thus draw the conclusion that vector bundle
$V_{(1,1,1,1,2,2;4,4)}$ is stable against a class of subsheaves
that has can be similarly described as $V$ itself.

\section*{Appendix B On finiteness of number of solutions to monad construction of vector bundles}

In this section, following \cite{LO}, we prove there are only
finite number of solutions to the sets of holomorphic vector
bundles which are described by monad construction and satisfy
anomaly cancellation condition on a Calabi-Yau manifold described
by intersecting hyperplanes in weighted projective space.

The rank-n vector bundle V is parameterized by two sets of
positive integers $(n_i;m_j) (i=1, \cdots, M+n; j=1, \cdots, M)$
constrained by
\begin{equation}
\sum_{i=0}^{n+M}n_i=\sum_{j=1}^M m_j, \qquad
\sum_{i=0}^{n+M}n_i^2+B =\sum_{j=1}^M m_j^2, \qquad m_j>n_i>0,
\qquad \forall i,j.
\end{equation}
where constant B is defined in (\ref{QValue}) and fixed by the
second Chern class of the Calabi-Yau manifold. We define $$ S
\equiv \sum_{i=0}^{n+M}n_i=\sum_{j=1}^M m_j $$ for convenience.

We first prove an inequality which will be essential for the
proof, \[ B\geq S. \] The proof is as follows:
\begin{eqnarray*}
B &\equiv& \sum_{j=1}^M m_j^2 - \sum_{i=0}^{n+M}n_i^2 \\
  &\geq & (n_{max}+1)\cdot \sum_{j=1}^M m_j-\sum_{i=0}^{n+M}n_i^2 \\
  &=& S+ \sum_{i=0}^{n+M} n_{max}n_i - \sum_{i=0}^{n+M}n_i^2 \\
  &\geq& S.
\end{eqnarray*}

Now we proceed in the following steps:

(1) n is a finite positive integer. The connection on the
holomorphic vector bundle takes value in a subgroup $G$ of $E_8$,
whose rank is finite. Further we consider the vector bundle $V$
being associated with the vector representation of $G$, whose rank
n is also finite.

(2) M is finite, actually, $M\leq\min\{\frac{B}{2}, B-n \}.$ The
proof is as follows: from $B\geq S$, one has
\[ B\geq \sum_{j=1}^M m_j \geq 2M, \]
as  $m_j>n_i>0$ so $m_i\geq 2$. This gives $M\leq \frac{B}{2}$. A
similar consideration for $S=\sum_{i=0}^{n+M}n_i$ produces the
other inequality.

(3) The maximal value $m_{max}$ of $m_j$ is finite, $m_{max}\leq
B-2(M-1)$. The proof is as follows,
\[ B\geq S = \sum_{j=1}^M m_j \geq m_{max} +2(M-1). \]

(4) The maximal value of $n_i$ is also finite. This is a simple
result of constraint $m_j>n_i>0$.

Thus we see that on a fixed Calabi-Yau manifold with fixed B, all
the parameters $(n, M, m_i, n_j)$ are constraint by certain
maximal values. Thus the number of solutions to the above
equations are finite.


\end{document}